\documentclass[usenatbib]{raa}            
\usepackage{graphicx,times}             
\usepackage{epstopdf}
\usepackage{lscape}
\usepackage{pdflscape}
\usepackage{colortbl}
\usepackage{amsmath}
\usepackage{textcomp}
\usepackage{multicol}
\usepackage{longtable}
\usepackage[figuresright]{rotating}
\usepackage{natbib}
\usepackage[colorlinks,  linkcolor=green,  anchorcolor=red,  citecolor=blue]{hyperref}
\begin{document}

   \title{Galaxy Properties Derived with Spectral Energy Distribution Fitting in the Hawaii-Hubble Deep Field-North 
}

   \volnopage{Vol.0 (20xx) No.0, 000--000}      
   \setcounter{page}{1}          

   \author{F. Y. Gao
      \inst{1,2}
   \and J. Y. Li
      \inst{1,2}
   \and Y. Q. Xue 
      \inst{1,2}
   }

   \institute{1. CAS Key Laboratory for Research in Galaxies and Cosmology, Department of Astronomy, University of Science and Technology of China, Hefei 230026, China; {\it fygao@mail.ustc.edu.cn, lijunyao@mail.ustc.edu.cn, xuey@ustc.edu.cn}\\
2. School of Astronomy and Space Science, University of Science and Technology of China, Hefei 230026, China\\
\vs\no
   {\small Received~~20xx month day; accepted~~20xx~~month day}}

\abstract{We compile multi-wavelength data from ultraviolet to infrared (IR) bands as well as redshift and source-type information for a large sample of 178,341 sources in the Hawaii-Hubble Deep Field-North field. 
A total of 145,635 sources among the full sample are classified/treated as galaxies and have redshift information available.
We derive physical properties for these sources utilizing the spectral energy distribution fitting code CIGALE that is based on Bayesian analysis.
Through various consistency and robustness check, we find that our stellar-mass and star-formation rate (SFR) estimates are reliable,
which is mainly due to two facts.
First, we adopt the most updated and accurate redshifts and point spread function-matched photometry; and second, 
we make sensible parameter choices with the CIGALE code and take into account influences of mid-IR/far-IR data, star-formation history models, and AGN contribution.
We release our catalog of galaxy properties publicly (including, e.g., redshift, stellar mass, SFR, age, metallicity, dust attenuation),
which is the largest of its kind in this field and should facilitate future relevant studies on formation and evolution of galaxies.    
\keywords{catalogs---galaxies: fundamental parameters---galaxies: star formation}
}

   \authorrunning{F. Y. Gao, J. Y. Li \& Y. Q. Xue}            
   \titlerunning{Galaxy Properties in H-HDF-N }  

   \maketitle

%

\section{Introduction}\label{sec:int}

\noindent Galaxies, where stars are born and die, are gravitationally bounded components constituting our universe. 
Tracing galaxy formation and evolution has been a major focus in many large sky surveys, such as the Cosmic Assembly Near-infrared Deep Extragalactic Legacy Survey (CANDELS; \citealt{Gro,Koe}) and the Galaxy and Mass Assembly survey (GAMA; \citealt{driver11,liske}). 
Understanding when and how galaxies form is key to deciphering the history and future of our universe.
Galaxy properties, such as the stellar mass ($M_*$), star-formation rate (SFR), intrinsic luminosity, and dust attenuation, are crucial in investigating many aspects of galaxies, e.g., the differences between active galactic nuclei (AGNs) and normal galaxies (see, e.g., \citealt{Net15} for a review), evolution of cosmic SFR (e.g., \citealt{MD15}), galaxy stellar-mass function (e.g., \citealt{baldry}), luminosity function (e.g., \citealt{Aird,Fin}), and hot dust-obscured galaxies (e.g., \citealt{Ass15}).

However, the big challenge for a survey is to identify sources as galaxies and carry out detailed follow-up spectroscopic and photometric research on them, especially for those with faint fluxes (e.g., \citealt{Bar08}). 
Due to flux-limit constraints of a survey, it is incredibly difficult to investigate galaxies that are in or close to the era of cosmic reionization from where it took over ten billion years for photons to reach us. 
That is why many epochs of the universe still remain largely unknown.
Despite of the difficulties mentioned above, many efforts have been dedicated to deriving physical properties of these very faint sources.

Thankfully, we have now access to enormous galaxy data owing to the advancement of large astronomical facilities and the devotion of astronomers around the world. 
Together, ground-based observatories and telescope arrays such as VLT, Keck, VLA, and ALMA, 
space observatories such as {\it HST}, \emph{Herschel}, {\it Chandra}, {\it NuSTAR}, {\it XMM-Newton}, and {\it Fermi}, 
and large observational projects lasting for decades such as the Sloan Digital Sky Survey (SDSS; \citealt{sdss}) and the 7~Ms {\it Chandra} Deep Field-south survey (CDF-S; \citealt{Xue11,Luo17,xuer}), 
cover a wide range of electromagnetic spectrum and build up a mass of multi-wavelength data. 
Moreover, many researchers have contributed dramatically in creating basic templates, 
e.g., the stellar population models (e.g., \citealt{BC03,M05}), dust attenuation curves (e.g., \citealt{Cal94,Cal00}), and dust emission templates in infrared (IR) bands (e.g., \citealt{CE01,DH02}). 
Given rich spectroscopic and photometric information that has become available recently, it is convenient to derive galaxy properties by fitting their observed spectral energy distributions (SEDs) with model templates. 
Compared with other methods that constrain one single parameter based on one single feature (e.g., deriving SFRs via the strengths of H$\alpha$ line), the approach of SED fitting can provide estimates of a number of parameters simultaneously (e.g., \citealt{Wal11}).

Galaxy SED fitting is now a widely-used technique that has gained popularity in recent decades. 
It has been proven that this method performs well in deriving a range of galaxy properties (e.g., redshift, $M_*$, SFR, dust mass, and metallicity) with high accuracy, 
as it calculates and optimizes these parameters self-consistently and simultaneously by taking into account how they influence each other and co-evolve (e.g., \citealt{Wal11,conroy,wright}).
However, there are some issues with SED fitting,
such as finer details of dust properties and dust-star geometries remaining to be investigated,
problems in estimating the age of the oldest stars,
as well as likely degeneracies between parameters (e.g., \citealt{Wal11}). 
Further improvements in SED fitting would be expected once a better understanding of these existing issues is obtained.

Despite of the wide use of SED fitting in deriving galaxy properties in tons of works, only a limited number of source catalogs in some sky areas that contain abundant information of galaxy properties (e.g., redshift, multi-wavelength photometry, $M_*$, and SFR) have been published. 
For instance, \cite{gal} presented an ultraviolet (UV) to mid-IR (MIR) catalog in the UKIRT Infrared Deep Sky Survey (UKIDSS) Ultra-Deep Survey field, which contains 35,932 {\it HST}/F160W-selected sources in a sky area of 201.7~arcmin$^2$ and provides their SEDs. 
\cite{mendel} provided a catalog of bulge, disk, and total stellar-mass estimates for SDSS DR7 that consists of almost 660,000 galaxies. 
They argued that, various SED modeling assumptions may result in an additional 60\% systematic uncertainty in mass estimation. \cite{S14} released a photometric catalog consisting of more than 200,000 sources in five fields, which also includes stellar masses derived from the FAST code (\citealt{Kriek09}). 
More recently, \cite{S15} combined the efforts from ten different teams, who derived galaxy properties using the same photometry and redshifts with several different codes for the GOODS-S and UDS fields. They found that, the stellar-mass estimates were largely affected by the choice of the stellar isochrone library. 
\cite{chang} combined the SDSS and WISE photometry for the full SDSS spectroscopic galaxy sample to create SEDs for a large and comprehensive catalog of 858,365 sources. They obtained the best-fit results using the MAGPHYS code (\citealt{MAGPHYS08}). 

In this paper, we compile multi-wavelength data from UV to IR in the Hawaii-Hubble Deep Field-North (H-HDF-N; \citealt{C04,Y14}), which is a 0.4 {deg}$^2$ area centering around the GOODS-N (\citealt{G04}) and {\it Chandra} Deep Field-North (CDF-N; \citealt{A03,Xue16,xuer}) fields. 
After carefully cross-matching between several catalogs that contain high-quality photometry and taking advantage of the SED-fitting technique, we produce a source property catalog that includes 145,635 galaxies, which is the largest of its kind in this field. 
After performing various robustness check, we find that our stellar-mass and SFR estimates are reliable.
We release our catalog publicly in order to facilitate future relevant studies.

This paper is structured as follows. Section~\ref{sec:data} lists the data cube and the pre-process before the SED fitting procedure. Section~\ref{sec:sed} describes the fitting parameters and their space. Comparisons between our results and others as well as robustness check are shown in Section~\ref{sec:com}. In Section~\ref{sec:discussion} we discuss the factors that can affect our results. The final catalog is presented in Section~\ref{sec:cat}. Finally, a brief summary of this work is included in Section~\ref{sec:sum}. Throughout this paper, all magnitudes are quoted in the
AB system, and we assume a cosmology with $H_0=70$~km~s$^{-1}$~Mpc$^{-1}$, $\Omega_M=0.27$, and $\Omega_{\Lambda}=0.73$.

\section{Data Collection and Compilation}\label{sec:data}
\subsection{Base Source Catalogs and Cross-matching Results}

We choose three catalogs as our base source catalogs: the photometric-redshift catalog in the \hbox{H-HDF-N} (\citealt{Y14}; hereafter Y14) that extracted point-spread function (PSF)-matched photometry in 15~bands and derived photometric redshifts using the EAzY code (\citealt{Brammer08}); the CANDELS/3D-{\it HST} catalog in the GOODS-N field (\citealt{S14}; hereafter S14) that presented photometric redshifts determined using the EAzY code and also galaxy properties derived from the FAST code; and the ultra-deep $K$s-band catalog in the GOODS-N field (\citealt{Wang10}). The details of each base catalog are summarized in Table~\ref{table:t1}.

\begin{table}
\begin{center}
\caption[]{Base Catalogs and Cross-matching Results}\label{table:t1}
\begin{tabular}{cccccc}
\hline\noalign{\smallskip}
Base catalog & Area (deg$^2$) & Source number & Matching radius & False-match rate & Matched sources \\
\hline\noalign{\smallskip}
 Y14 & 0.40 & 131,678 &--- &---&--- \\
 S14 & 0.05 & 38,279&$0.5\arcsec$ & 5.6\% &17,014\\
 \cite{Wang10} & 0.25 &94,951& $0.5\arcsec$ &3.4\%&69,553\\
 \noalign{\smallskip}\hline 
Combined catalog &---& 178,341 &--- &---&---  \\
\noalign{\smallskip}\hline
\end{tabular}
\end{center}
\end{table}

Aiming at creating a most complete combined source catalog, we cross-match three base catalogs one by one in the order mentioned above to reach a maximum source number. 
We adopt 0.5\arcsec~as the matching radius. 
To account for the small non-uniformity in the astrometry between different catalogs, we match each catalog twice so as to increase the matching accuracy. For the first matching, we calculate the mean positional deviation between two individual catalogs, and correct the derived mean deviation (i.e., mean differences in RA and DEC between the matched sources) to each source of one catalog. 
We then perform the second matching using the updated astrometry. During each concatenation of the three base catalogs, only the coordinates of the unmatched sources of the second catalog will be appended to the coordinate list of the first catalog.
By this means we obtain a combined source catalog that is composed of 178,341 sources, i.e., $131,678+(38,279-17,014)+(94,951-69,553)=178,341$ (see Table~\ref{table:t1}).  
While matching the remaining multi-wavelength catalogs and redshift catalogs to this combined source catalog, again, using a matching radius of 0.5\arcsec~(see Section~\ref{subsec:redshift_sed}), the unmatched parts are discarded, i.e., the source number of 178,341 remains unchanged for this operation.

In each matching step, we shift one catalog in 8 directions (i.e., 4 axial directions in terms of right ascension and declination and 4 diagonal directions) with a 30\arcsec~displacement and then re-correlate the sources in order to estimate the false-match rate (i.e., the mean matched source number divided by the original non-shifted matched source number).
After the first-step astrometry correction, the false-match rate drops and the matched fraction rises in the second step. 
The final average false-match rates are in the range of 3.1\% to 5.6\%, with a typical value around 5\% (see Table~\ref{table:t1} for example). The lowest rate occurs when matching the spectroscopic-redshift catalog of \cite{Wir04} to the combined source catalog, due to the limited number of sources in \cite{Wir04}; while the largest rate results from matching the S14 base catalog to Y14 base catalog, due to the great depth (hence large source surface density) of the S14 catalog. Overall, these false-match rates are small and will not significantly influence our results.

\subsection{Redshift Information and Broadband SED Construction}\label{subsec:redshift_sed}

The next step is to collect source-type and redshift information for the 178,341 sources in the combined sample. 
We compile spectroscopic and photometric type classifications from a number of references and their relevant information is listed in Table~\ref{table:t2}. 
We determine the type of one source according to the following selection order of preference: spectroscopic type $>$ S14 photometric type $>$ Y14 photometric type. For X-ray-detected sources, we adopt the final classifications presented in \cite{Xue16} CDF-N main and supplementary catalogs. 
A total of 152,961 sources out of the 178,341 sources have type information, and the remaining sources have no such information in their respective original catalogs.

\begin{table}
\begin{center}
\caption[]{Source-type Information}\label{table:t2}
\begin{tabular}{ccccccccc}
\hline\noalign{\smallskip}
Index &Reference & Matched Number &Star& Galaxy&Undistinguished&White Dwarf&No Info& X-ray AGN\\
\hline\noalign{\smallskip}
&&Spectroscopic&&&Catalog&&&\\
\hline\noalign{\smallskip}
 1&\cite{Bar08}&2606&192&2414&0&0&0&0\\
2&\cite{Wir04}&2606&10&2596&0&0&0&0\\
3&\cite{Cowie}&66&10&56&0&0&0&0\\
4&\cite{Coo11}&81&2&79&0&0&0&0\\
5&S14&82&0&0&0&0&82&0\\
6&Y14&126&0&0&0&0&126&0\\
 \noalign{\smallskip}\hline
&Subtotal:&5,567&214&5,145&0&0&208&0\\
\hline\noalign{\smallskip}
&&Photometric&&&Catalog&&&\\
\hline\noalign{\smallskip}
7&S14&32,754&138&7655&24,961$^a$&0&0&0\\
8&Y14&113,944&4,310&109,602&0&24&0&8\\
 \noalign{\smallskip}\hline
&Subtotal:&146,698&4,448&117,257&24,961&24&0&8\\
 \noalign{\smallskip}\hline
&&X-ray&&&Catalog&&&\\
\hline\noalign{\smallskip}
9&\cite{Xue16}$^{b}$&627&16&75&0&0&0&536\\
10&\cite{Xue16}$^{b}$&69&0&37&0&0&0&32\\
 \noalign{\smallskip}\hline
& Subtotal:&696&16&112&0&0&0&568 \\
 \noalign{\smallskip}\hline
&Total:&152,961$^{c}$&4,678&122,514&24,961&24&208&576\\
 \noalign{\smallskip}\hline
\end{tabular}
\end{center}
\tablecomments{0.96\textwidth}{(a) These sources in S14 are too faint to be classified with confidence and thus are assigned a specific type of ``Undistinguished" (labeled as type=2 in our final catalog; see Section~\ref{sec:cat}), which we retain for SED fitting once they have non-zero redshifts;\\
	(b) Type classifications are provided by the \cite{Xue16} CDF-N main and supplementary catalogs, respectively;\\
(c) This number is smaller than 178,341, due to that some catalogs do not provide type classifications.}
\end{table}

\begin{table}
\begin{center}
\caption[]{Redshift Information$^a$}\label{table:t3}
\begin{tabular}{ccccccc}
\hline\noalign{\smallskip}
Index &Reference & Matched Number& Galaxy&Undistinguished&No Info& X-ray AGN\\
\hline\noalign{\smallskip}
&&Spectroscopic&Catalog&&&\\
\hline\noalign{\smallskip}
1&\cite{Bar08}&2,414&2,414&0&0&0\\
2&\cite{Wir04}&141&141&0&0&0\\
3&\cite{Cowie}&56&56&0&0&0\\
4&\cite{Coo11}&79&79&0&0&0\\
5&S14&79&0&0&79&0\\
6&Y14&118&0&0&118&0\\
\noalign{\smallskip}\hline
&Subtotal:&2,887&2,690&0&197&0\\
\hline\noalign{\smallskip}
&&Photometric&Catalog&&&\\
\hline\noalign{\smallskip}
7&S14&32,458&7,552&24,906&0&0\\
8&Y14&109,610&109,602&0&0&8\\
\hline\noalign{\smallskip}
&Subtotal:&142,068&117,154&24,906&0&8\\
\noalign{\smallskip}\hline
&&X-ray&Catalog&&&\\
\hline\noalign{\smallskip}
9&\cite{Xue16}$^{b}$&611&75&0&0&536\\
10&\cite{Xue16}$^{b}$&69&37&0&0&32\\
 \noalign{\smallskip}\hline
& Subtotal:&680&112&0&0&568\\
 \noalign{\smallskip}\hline
&Total:&145,635&119,956&24,906&197&576\\
 \noalign{\smallskip}\hline
\end{tabular}
\end{center}
\tablecomments{0.86\textwidth}{(a) Stars and sources without redshift information are not included for SED fitting. \\
(b) Redshifts are provided by the \cite{Xue16} CDF-N main and supplementary catalogs, respectively.}
\end{table}

We follow a preference order of spectroscopic redshift $>$ S14 photometric redshift $>$ Y14 photometric redshift when choosing the redshift for each source. We use the 10,552 sources that are classified as galaxies and have photometric redshifts from both S14 and Y14 to evaluate the consistency between redshift estimates. The two sets of photometric redshifts are overall in reasonable agreement, with the normalized median absolute deviation $\sigma_{\rm NMAD}=0.058$ and an outlier (defined as $|\Delta z|/(1+z_{\rm Y14})>0.15$ where $\Delta z=z_{\rm Y14}-z_{\rm S14}$) fraction of 13.3\%. The difference between the photometric-redshift estimates may be due to the differences in the adopted galaxy templates, photometry, and other fitting details.

Since we focus on deriving galaxy properties, for sources that have been classified as stars, their redshifts are fixed to 0 and not included for further SED fitting.
In some references, a fraction of sources are not explicitly classified as stars or galaxies. 
For example, a total of 24,961 sources adopted from S14 are too faint to be classified with confidence, they are labeled as undistinguished type (see Table~\ref{table:t2}), and are retained in our SED fitting routine as long as they have non-zero redshifts, i.e., we treat them as galaxies.

\begin{figure}
 \centering 
    \includegraphics[width=\textwidth,angle=0]{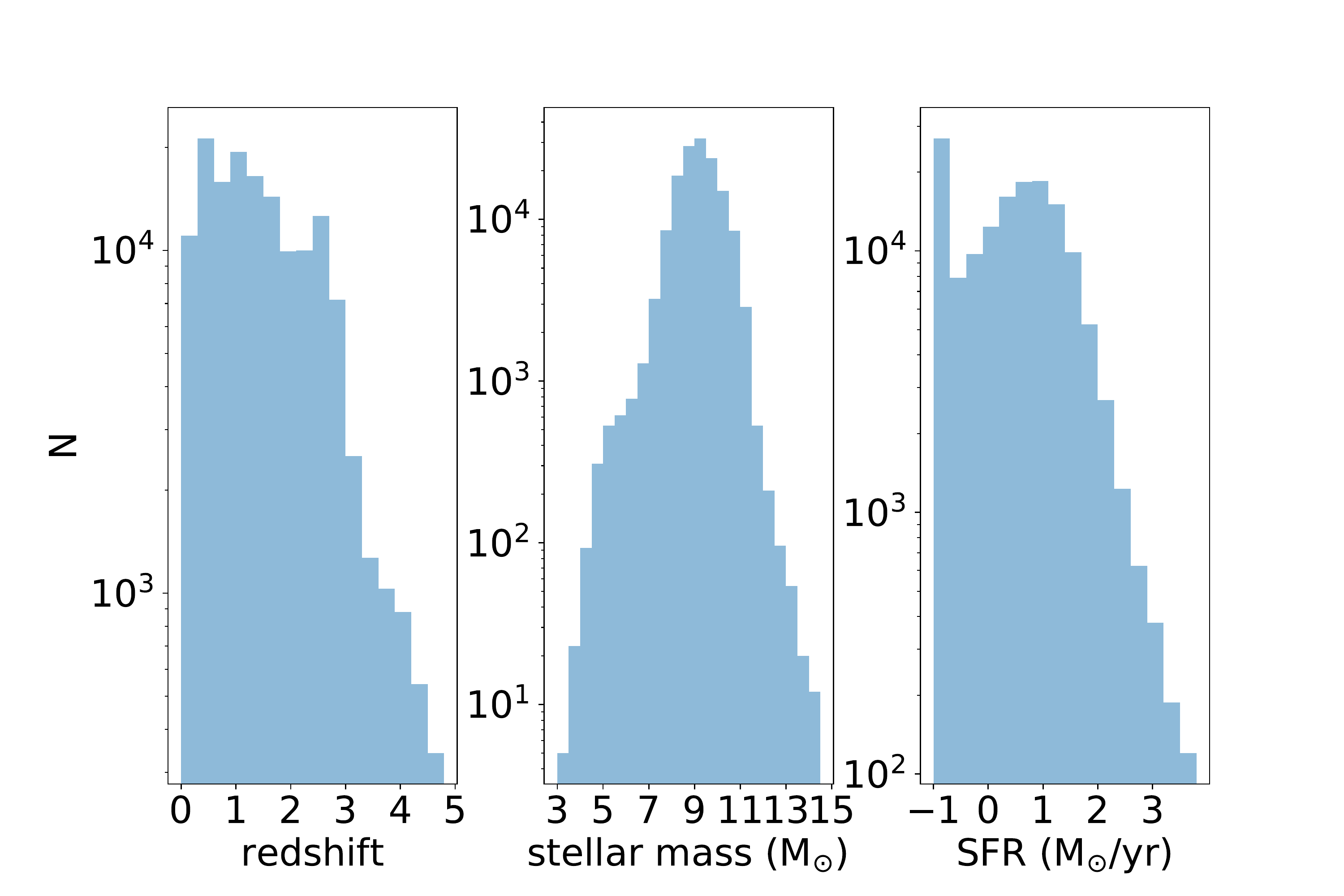}
    \caption{Distributions of redshift, stellar mass (logarithmic scale), and SFR (logarithmic scale; values of $<$0.1 set to 0.1 $M_\odot$ yr$^{-1}$) for the final galaxy sample of 145,635 sources. The first quartile, median and third quartile errors are 0.168, 0.235, 0.334 dex for stellar mass and 0.346, 0.420, 0.553 dex for SFR, respectively.}
    \label{fig:result}
\end{figure}

After these procedures, we obtain the final galaxy sample of 145,635 galaxies (or treated as galaxies)
that have redshift information available (98\% being photometric redshifts) out of the 178,341 sources from the original sample. The redshift information is listed in Table~\ref{table:t3}. The redshift distribution of the final galaxy sample is shown in the left panel of Figure~\ref{fig:result}. Our sample covers a broad redshift span ranging from $z \sim 0$ to $z \sim 5$. It also contains a significant population at $z \sim 2$, which is well known as the peak epoch of star-formation activity. Therefore, our sample is ideal for investigating stellar-mass assembly and evolution of galaxies.

After the final galaxy sample is secured, we start to construct the broadband SED ranging from UV to IR (i.e., $\lambda  \le 8 \rm \mu m$ in this work) for each source. We list the information of the photometric bands we adopt in Table \ref{table:t4}. More details are referred to Table 1 of Y14, Table 4 of \cite{Wang10}, Tables 2--6 of \cite{Ash13}, and Table 6 of S14 (we exclude IRAC1, IRAC2, Subaru B, V, z', i' and r' filters according to the known issues listed in the 3D-HST website). All the photometric data used in SED fitting, when necessary, are corrected to be the total fluxes for aperture effect. In addition, for sources with only flux errors being provided in the original references (which means no detection), the flux errors are treated as 1$\sigma$ upper limits on fluxes (i.e., error$\pm$error).

Most of our multi-wavelength data are from Y14, which provides PSF-matched photometry from a number of optical-to-IR images. For one band that has two photometric data points available given by different references, we make the following arrangement of photometry: (a) If one of the two origins is a PSF-matched catalog, we assign this photometry a higher priority with the other being supplementary. During supplementation, we select a neighboring band as the reference band. Only the supplementary photometry that lies within 1 dex (in terms of flux level) of the reference band is retained. (b) If two sets of photometry are both from PSF-matched catalogs or aperture-corrected catalogs, we perform a consistency check: if the absolute flux difference is smaller than the sum of two flux errors, the mean value is adopted; otherwise, we keep the one whose flux is closer to the reference band. (c) When one certain band has more than two origins, we select one or two as the highest priority according to the reliability of their photometry, and then follow the aforementioned procedures.

\begin{table}
\begin{center}
\caption[]{List of Multi-wavelength Data$^a$}\label{table:t4}
\begin{tabular}{ccccc|ccc}
\hline\noalign{\smallskip}
Band&Depth&PSF Size&Zero Point&Reference&Depth&Aperture&Reference\\  
& (mag) & & (mag) & & (mag) \\
\hline\noalign{\smallskip}
$U$&26.3&1.26\arcsec &31.369&Y14&26.4&1.5\arcsec&S14\\
$B$&26.3&1.13\arcsec &31.136&Y14&---&---&---\\
$V$&25.8&1.56\arcsec &34.707&Y14&---&---&---\\
$R$&26.0&1.60\arcsec &34.676&Y14&---&---&---\\
$I$&25.1&1.08\arcsec &33.481&Y14&---&---&---\\
$z$&24.9&1.08\arcsec &33.946&Y14&---&---&---\\
$J^{b}$&24.5&1.11\arcsec &23.900&Y14&---&---&---\\
$J^b$&---&---&---&---&25.0&1.0\arcsec&S14\\
$H$&22.9&1.32\arcsec &23.900&Y14&24.3&1.0\arcsec&S14\\
$HK^{'}$&22.3&1.20\arcsec &30.132&Y14&---&---&---\\
$Ks^{b}$&23.7&1.08\arcsec &23.900&Y14&0.18$\mu$Jy$^{d}$&5\arcsec &\cite{Wang10}\\
$Ks^b$&---&---&---&---&24.7&1.0\arcsec&S14\\
3.6$\mu$m$^{c}$&25.1/24.5&2.53\arcsec/2.40\arcsec &21.581/21.581&Y14&0.11$\mu$Jy$^{d}$&4-6\arcsec &\cite{Wang10}\\
4.5$\mu$m$^{c}$&24.6/24.2&2.53\arcsec /2.43\arcsec &21.581/21.581&Y14&0.12$\mu$Jy$^{d}$&4-6\arcsec &\cite{Wang10}\\
5.8$\mu$m&22.6&2.96\arcsec &21.581&Y14&0.42$\mu$Jy$^{d}$&4-6\arcsec &\cite{Wang10}\\
&&&&&22.8&3.0\arcsec&S14\\
8.0$\mu$m&22.7&3.24\arcsec &21.581&Y14&0.42$\mu$Jy$^{d}$&4-6\arcsec &\cite{Wang10}\\
&&&&&22.7&3.0\arcsec&S14\\
F606W&---&---&---&---&27.4&0.7\arcsec&S14\\
F125W&---&---&---&---&26.7&0.7\arcsec&S14\\
F140W&---&---&---&---&25.9&0.7\arcsec&S14\\
F160W&---&---&---&---&26.1&0.7\arcsec&S14\\
F435W&---&---&---&---&27.1&0.7\arcsec&S14\\
F775W&---&---&---&---&26.9&0.7\arcsec&S14\\
F860LP&---&---&---&---&26.7&0.7\arcsec&S14\\
G&---&---&---&---&26.3&1.2\arcsec&S14\\
Rs&---&---&---&---&25.6&1.2\arcsec&S14\\
 \noalign{\smallskip}\hline  
\end{tabular}
\end{center}
\tablecomments{0.86\textwidth}{(a) The left and right portions of this table are for the PSF-matched and aperture photometry, respectively.\\ (b) These are two slightly different filters from two telescopes with the same name.\\	
(c) These two photometric bands in Y14 have two origins.\\
(d) These depths (in units of $\mu$Jy) from \cite{Wang10} are median 1$\sigma$ errors among sources detected at $\ge3\sigma$, unlike other depths (in units of magnitude) quoted at 5$\sigma$.
}
\end{table} 

\section{SED Fitting Parameters}\label{sec:sed}

To obtain crucial galaxy properties of our sample, we carry out multi-wavelength SED fitting using the PCIGALE code to fit the broadband SED as constructed in Section \ref{subsec:redshift_sed}. PCIGALE is a python version of CIGALE (Code Investigating GALaxy Evolution)\footnote{See https://cigale.lam.fr/.} whose algorithm was first written by \cite{Bur05} and further developed by \cite{Noll}. Based on Bayesian analysis, CIGALE can reliably derive a variety of galaxy parameters, such as $M_*$, stellar age, SFR, and dust attenuation, which are vital to understanding formation and evolution of galaxies. 

In order to fit the observed SEDs, first we have to build a template library that contains various galaxy SEDs with different parameters. The galaxy SEDs are generated from the combination of several modules, such as the stellar population model \citep[][hereafter BC03]{BC03}, the dust attenuation curve \citep{Cal00}, the IR emission by dust that originates from UV/optical photons being absorbed and re-radiated in longer wavelengths \citep{DL07,Dale}, as well as as the AGN radiation \citep{Fritz}. These modules are all incorporated into CIGALE that are convenient for us to invoke. Below we describe the modules used in the fitting routine and the adopted parameter space.

Star-formation history (SFH) is one of the most important component that determines the galaxy growth history and has significant impact on the derived galaxy properties. In a real galaxy the SFH can be very complicated and difficult to depict. In the attempt of reproducing the real SFH using simple functions, several scenarios have been proposed. For example, the exponentially declining model assumes that the SFR decreases exponentially with a characteristic timescale $\tau$ (i.e., the 1$\tau$-dec model); the 2$\tau$-dec model adds a late starburst to the 1$\tau$-dec model; and the so-called delayed-$\tau$ model represents an earlier rising and then declining SFR evolution (e.g., $\mathrm{SFR}(t)\propto t/\tau^2 \cdot {\rm exp}(-t/\tau)$; see \cite{S15}).          
We choose the delayed-$\tau$
model as our SFH since \cite{ciesla} demonstrated that this SFH provides better estimates of galaxy properties compared with both observations and hydrodynamical simulations (see Section \ref{subsec:sfh} for relevant discussions). Moreover, this model has fewer parameters, which effectively reduces degeneracy. We allow both $\tau$ and the oldest stellar age $t$ to vary between 1 and 10000 Myr.

The BC03 simple stellar population (SSP) synthesis model is used to generate the emission from the stellar population by assuming a Chabrier initial mass function \citep[IMF;][]{Cha03}. In terms of metallicity, CIGALE code provides several options at choice, ranging from 0.02 to 2.5 times solar metallicity.

The attenuation of the stellar emission due to absorption and scatter by dust grains is described by extinction curves. We utilize the widely used Calzetti attenuation law \citep{Cal00} with the color excess E$(B-V)$ of the young population ranging from 0 to 1, while freezing the reduction factor for the color excess of the older population to a constant default value 0.44.

Dust re-radiates the absorbed UV and optical photons predominantly at longer wavelengths. CIGALE provides four sets of dust emission templates \citep{DH02,DL07,casey,Dale}. Given the time-consuming characteristic of the fitting procedure, we select the \cite{Dale} template set since it has fewer free parameters and models the AGN emission simultaneously. Considering the limited number of X-ray detected sources in our sample that are usually moderately luminous \cite[see][]{Xue10,Xue16}, we assume a low to moderate AGN contribution to the total galaxy SED, i.e., the AGN fraction varies between 0.0 and 0.6 (see Section \ref{subsec:agn} for relevant discussions). The heating intensity slope, $\alpha$, which describes the peak wavelength of the dust emission \citep[see Figure 5 in][]{Noll}, is allowed to vary among 1.5, 2.0, and 2.5. The full parameter space we adopt for each module is summarized in Table \ref{table:t5}. During the fittting, the CIGALE code itself discards unphysical solutions
where galaxies are older than the age of the Universe at their respective redshifts.

\begin{table}
\begin{center}
\caption[]{Parameter Space of the Delayed-$\tau$ Model}  \label{table:t5}
\begin{tabular}{ccc}
 \hline\noalign{\smallskip}
Module & Parameter & Value \\ 
 \hline\noalign{\smallskip}
    star-formation history&$\tau$&1--10000 Myr (9 steps)$^a$\\ 
    delayed-$\tau$ model&age&1--10000 Myr (19 steps)$^b$ \\ 
    \noalign{\smallskip}\hline
    simple stellar population & initial mass function & \cite{Cha03} \\ 
    BC03 & metallicity (solar) & 0.02, 0.2, 0.4, 1, 2.5 \\ 
    &separation age & 1, 5, 10, 50, 100, 500, 1000 Myr \\
    \noalign{\smallskip}\hline
    dust attenuation &E$(B-V)$ of young population & 0--1.0 (in steps of 0.1)\\
    \noalign{\smallskip}\hline
    dust emission \&  AGN contribution & AGN fraction &0.0--0.6 (in steps of 0.1) \\
    &heating intensity $\alpha$ & 1.5, 2, 2.5 \\
\noalign{\smallskip}\hline
\end{tabular}
\end{center}
\tablecomments{0.86\textwidth}{(a) The 9 values are 1, 5, 10, 50, 100, 500, 1000, 5000, and 10000 Myr.\\
(b) The 19 values are 1, 5, 10, 50, 100, 200, 400, 600, 800, 1000, 2000, 3000, 4000, 5000, 6000, 7000, 8000, 9000, and 10000 Myr.}
\end{table}

\section{SED Fitting Result and Reliability Check}\label{sec:com}

\begin{figure}[h]
    \includegraphics[width=\linewidth]{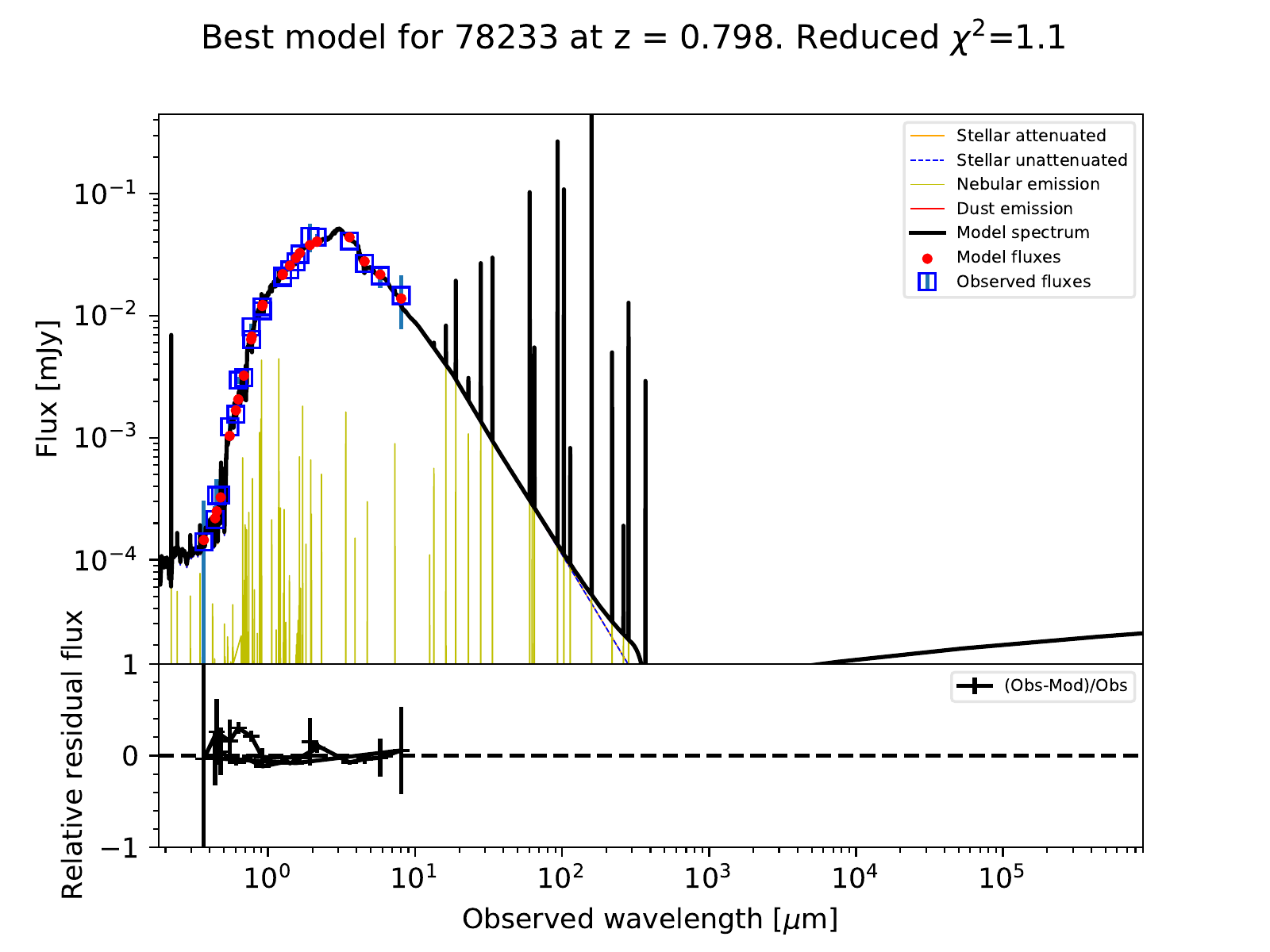}
    \caption{{\small Best-fit SED model for the source with id=78233 as provided by CIGALE.}}
\label{fig:bestsed}
\end{figure}

Figure \ref{fig:bestsed} presents an example of the best-fit SED given by CIGALE. 
The distributions of the derived stellar masses and SFRs are demonstrated in the middle and right panel of Figure \ref{fig:result}, respectively. 
For simplicity, we set the SFRs and/or their errors less than 0.1$\,M_{\odot}\,\rm yr^{-1}$ to 0.1$\,M_{\odot} \rm\,yr^{-1}$. 
The striking spike that denotes this considerably small SFR value in the right panel appears ubiquitous in galaxy property catalogs based on SED fitting 
\cite[see, e.g., results from ten groups in][]{S15}.  

In the remainder of this section, we check the reliability of our result and the consistency with previous works through various methods. Since the most concerned galaxy properties are stellar mass and SFR, we only consider these two parameters here, and also include the other parameters (i.e., stellar age, metallicity, and dust attenuation) in the final catalog (see Section~\ref{sec:cat}). 
In order to make a meaningful and fair comparison, 
we require our sources for comparison to have data (including upper limits) on $\ge10$ filters (unless specified otherwise) in order to ensure reasonable quality of SED fitting results. Furthermore, 
we require a redshift match between the common sources in the reference works (denoted as $z1$) and our work (denoted as $z2$), i.e., 
$|z_1-z_2|/(1+z_1)\le0.05$. 
Note that this is a relatively loose requirement, especially at high redshifts; different redshift values adopted in various works can lead to inconsistent SED fitting results.

\subsection{Stellar Mass Comparison}\label{subsec:mass}

\begin{figure}[h]
    \includegraphics[width=0.95\linewidth]{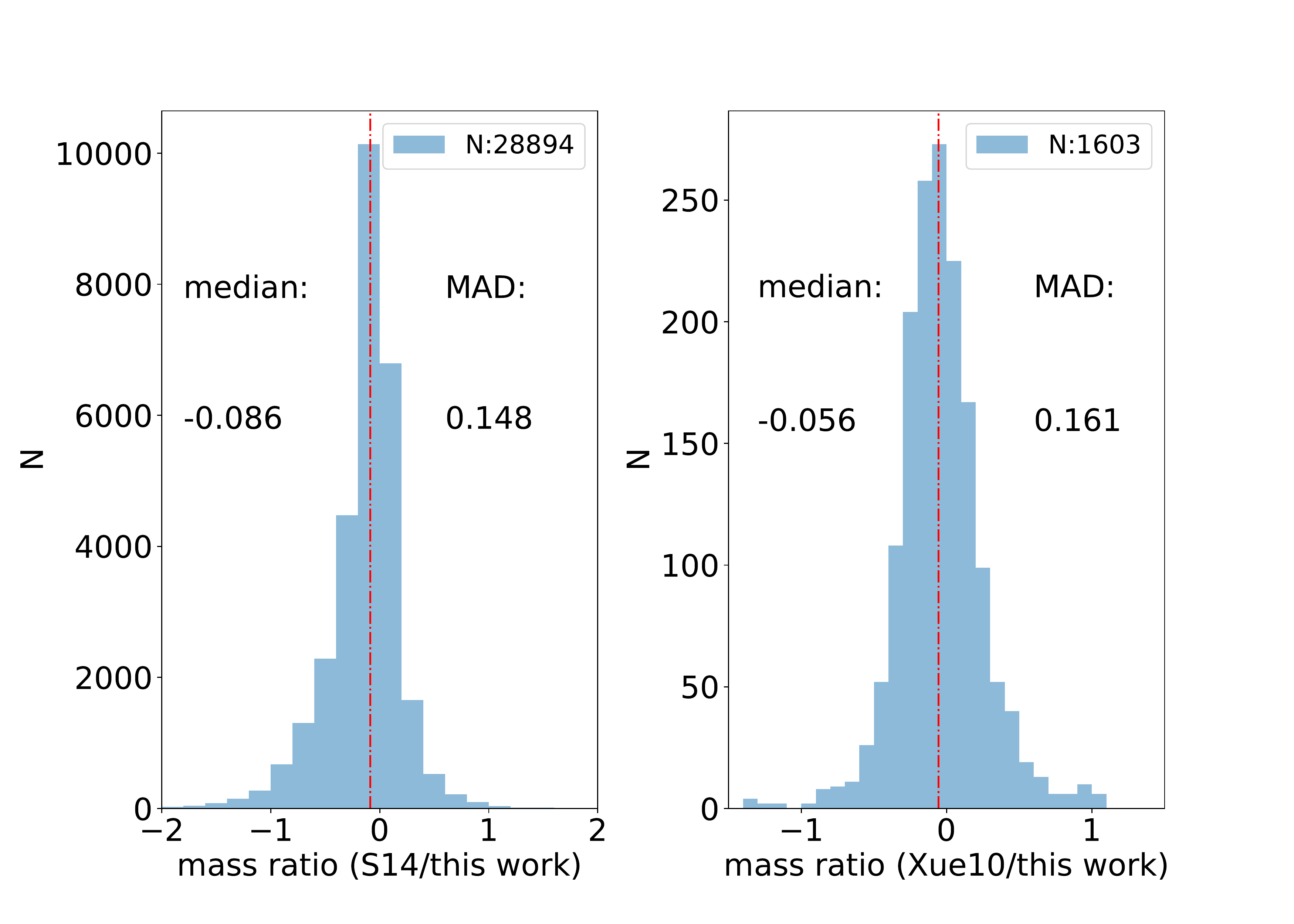} 
    \caption{{\small (Left) Histogram of the logarithmic stellar-mass ratio between S14 and our work, with the median value (shown as the dashed-dot line), MAD, and number of matched pairs annotated. (Right) Same as the left panel, but for the comparison between Xue10 and our work.}}
\label{fig:mass_ratio}
\end{figure}

We first compare our stellar mass estimation with S14 as shown in the left panel of Figure \ref{fig:mass_ratio}. 
S14 adopted the 1$\tau$-dec SFH, the BC03 stellar population synthesis model with a \cite{Cha03} IMF, and solar metalicity when deriving stellar mass using the FAST code. We note that, as S14 is one of the three base catalogs, our final catalog retains much photometry and redshift data from it. 
As a result, it is not surprising that stellar masses derived from our work are well consistent with theirs. 
Among the 28,894 redshift-matched pairs, the median value of the logarithmic stellar-mass ratio between S14 and our work is $-$0.086 dex, and the median absolute deviation (MAD) is 0.148 dex
(in subsequent comparisons, all ratios are quoted in the logarithmic scale and we omit the logarithmic term hereafter for brevity). 

We then conduct another comparison with \citet[][hereafter Xue10]{Xue10} that calculated the stellar mass based on the tight correlations between rest-frame optical colors and mass-to-light ratios.  
The result is shown in the right panel of Figure \ref{fig:mass_ratio}. Given that the formula used in Xue10 (i.e., their Equation 1) was adapted with a \cite{Kroupa} IMF, we divide the stellar mass in Xue10 by a factor of 1.06 according to the Equation 2 in \citet[][and references therein]{speagle}, in order to make their estimates adapt our adopted \cite{Cha03} IMF. 
We find a median stellar-mass ratio of $-0.056$ dex and MAD=0.161 dex for the 1603 redshift-matched pairs.

According to the above comparisons, it is clear that our stellar-mass estimation is in good agreement with that of previous works (i.e., median offsets $<$0.1 dex), 
albeit with reasonably small scatters (i.e., MAD$\sim$0.15 dex)
that arise from differences in the adopted photometry, redshifts, and derivation details.
Therefore, our SED fitting procedure is able to provide robust stellar-mass estimates.

\subsection{SFR Comparison} \label{subsec:sfr}

We then check the consistency of SFR estimates with previous works. In the first place, we compare our work with \cite{W14} (hereafter W14) that derived SFRs for S14 sources using empirical relations (see W14 for details) and with Xue10. The latter separated their sample into two groups: sources with \emph{Spitzer} MIPS 24 $\mu$m detection and without (denoted as upper limits). For the MIPS detected sources, they calculated SFRs using the empirical relation between SFR and UV+IR luminosity (in units of solar luminosity) as described by the following equation:
\begin{equation}
    {\rm SFR}\,(M_\odot \,{\rm yr}^{-1})=9.8 \times 10^{-11}(L_{\rm UV}/L_\odot+L_{\rm IR}/L_\odot).
\label{eq:sfr_uv_ir}
\end{equation}
The IR luminosity is extracted from the 24 $\rm \mu m$ flux -- $L_{\rm IR}$ correlation using the \cite{CE01} templates, while the UV luminosity is calculated from the best-fit SED (see Section 3.3 of Xue10 for details). 
For the MIPS undetected sources, they presented upper limits on SFRs again using Equation \ref{eq:sfr_uv_ir} and also provided SED fitting derived SFRs in their full unpublished catalog.

\begin{figure}
    \centering
    \includegraphics[width=\linewidth,angle=0]{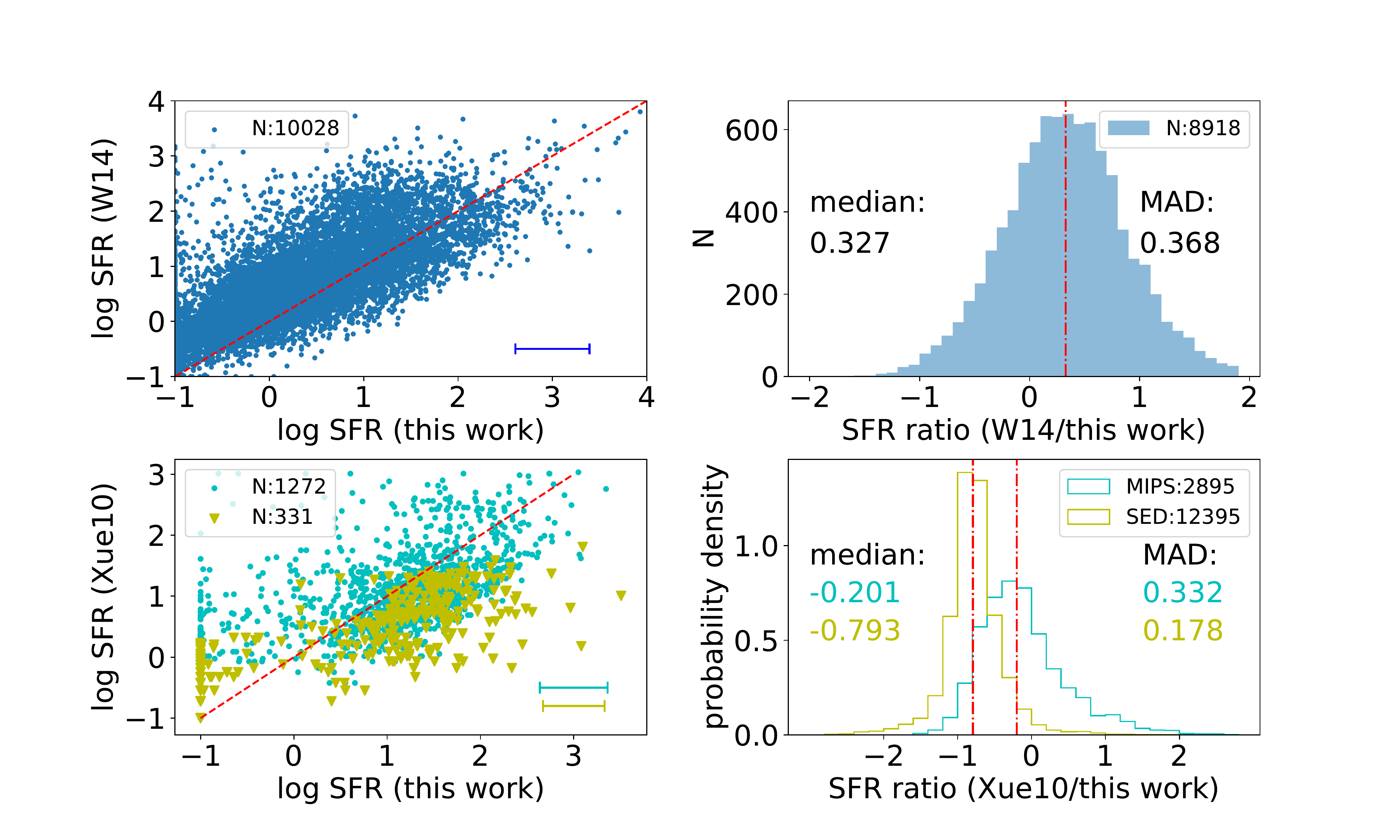} 
    \caption{\small (Top) SFR comparison with W14. We only show sources with better MIPS meausurements and exclude SFRs less than 0.1$M_\odot \,\mathrm{yr^{-1}}$ for upper right panel. (Bottom Left) SFR comparison with the Xue10 published catalog. The cyan dots and yellow triangles denote sources with (1272 sources) and without (331 sources with upper limits on SFRs) MIPS 24 $\mu$m detection in Xue10, respectively. The red dashed line represents the 1-to-1 relation. 
(Bottom Right) Histograms of the logarithmic SFR ratio between the Xue10 full unpublished catalog and this work. We exclude sources with SFRs less than 0.1$M_\odot \,\mathrm{yr^{-1}}$. The cyan and yellow colors represent 2895 sources with and 123,95 sources without MIPS 24 $\mu$m detection in Xue10, respectively; 
and the latter (i.e., yellow) corresponds to SED fitting derived SFRs in Xue10. From this figure on, the short (horizontal and/or vertical) segments in the panel corners indicate the respective median errors (when available) on the measured parameters (SFR here) multiplied by 2.
}
    \label{fig:sfr}
\end{figure}

The comparison result is shown in  Figure \ref{fig:sfr}. 
The cyan dots and yellow triangles represent sources with and without MIPS 24 $\mu$m detection in Xue10, respectively. 
Clearly, for sources matched with W14 (top panels) and Xue10 with 24 $\mu$m detection (bottom panels), there exists an over-density above/below the one-to-one line respectively, while for those without 24 $\rm \mu m$ detection (i.e., upper limits), our SFR estimates are systematically larger than Xue10. The median ratio of the W14 SFR estimates to ours reaches $\sim 0.3$ dex (see explanations below).
To increase the source number for comparison with Xue10, we then compare with the full unpublished catalog of Xue10 in the bottom-right  panel of Figure \ref{fig:sfr}. 
We also find deviations between Xue10 and our SFR estimates (e.g., a median offset of $-0.201$ dex and MAD=0.332 dex for the 2,895 matched pairs with MIPS detection).
However, we note that the 24$\mu$m photometry used in Xue10 was extracted from an unpublished catalog, the majority of which has relatively low signal-to-noise ratios. Furthermore, the photometric data used in Xue10 and our work are adopted from different catalogs and may have systematic differences. As mentioned in Y14, the more accurate PSF-matched photometry published in their work (and adopted by our work) has a 0.1 mag median offset and a 0.15 mag scatter compared with aperture-corrected photometry (as adopted in, e.g., Xue10). Therefore, the different photometric data and redshifts adopted in various works all have significant influences on SFR estimation.
In addition, we notice that the derived SFRs can still be poorly constrained and highly uncertain
even with the same photometry and redshifts. We compare the SFR estimates derived from ten groups in \cite{S15}, who utilized exactly the same data but different SED fitting parameters and codes, and find large scatters and deviations among results of each group.
Therefore, the SFR estimation is highly data and model dependent.

\begin{figure}
    \centering
    \includegraphics[width=\linewidth,angle=0]{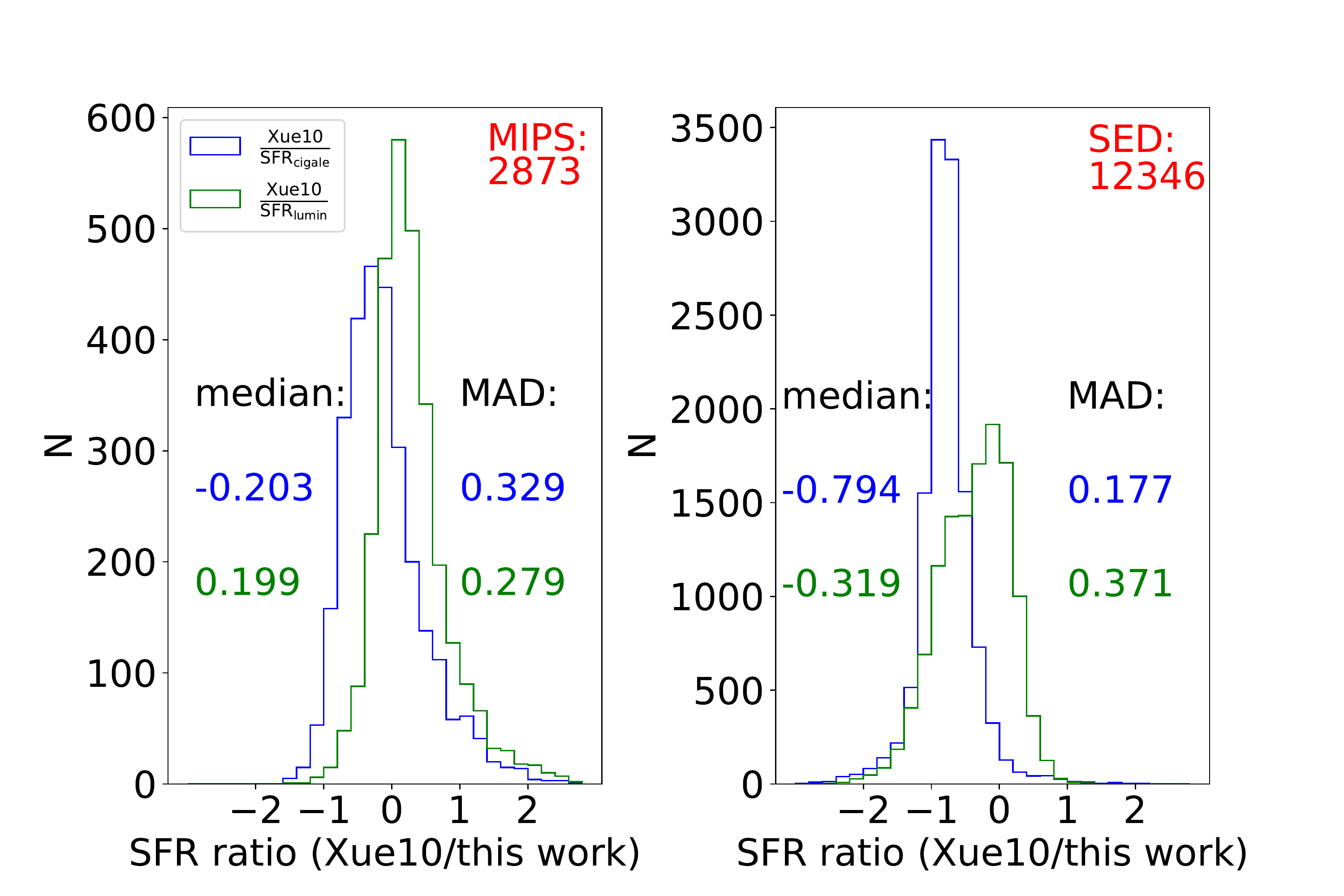} 
    \caption{\small (Left) Histograms of the logarithmic SFR ratio between Xue10 and our work for the 2873 sources with MIPS detection in Xue10. 
The blue (green) color is for the comparison between Xue10 SFR and our SFR$_{\rm cigale}$ (SFR$_{\rm lumin}$). (Right) Same as the left panel, but for the 12,346 sources without MIPS detection in Xue10. In both panels, sources with Xue10 SFR/SFR$_{\rm cigale}$/SFR$_{\rm lumin}$ $<$0.1 $M_\odot$ yr$^{-1}$ are not included.}
    \label{fig:cal1}
\end{figure}

Aiming at investigating whether the disagreement in SFR estimation is partly due to the different calculation methods adopted in Xue10 and our work, we also utilize Equation \ref{eq:sfr_uv_ir} (correcting for IMF difference) as in Xue10 to estimate SFRs and make another comparison. The $L_{\rm IR}$ is calculated by integrating the dust emission of the best-fit SED for each source in the rest-frame 8--1000 $\mu$m, while the $L_{\rm UV}$ is 
represented by 3.3 times the rest-frame 2800 $\rm \AA$ galaxy luminosity based on the best-fit SED (see Section 3.3 of Xue10). 
The comparison is shown in Figure \ref{fig:cal1}. The blue histograms are for the logarithmic ratios between Xue10 SFRs and our SFRs derived with CIGALE (denoted as SFR$_{\rm cigale}$), while the green histograms are for that between Xue10 SFRs and our SFRs derived using Equation \ref{eq:sfr_uv_ir} (denoted as SFR$_{\rm lumin}$).
It is obvious that when we adopt the same formula as in Xue10 to calculate the SFR, the results vary significantly and appear more consistent with those from Xue10 in some aspects but not the others: 
for the 2873 sources with MIPS detection, the median offset changes from $-0.203$ to 0.199 dex, and MAD reduces from 0.329 to 0.279 dex;
for the 12,346 sources without MIPS detection, the median offset reduces from $-0.794$ to $-0.319$ dex, but MAD increases from 0.177 to 0.371 dex.
This demonstrates that the SFR estimation is highly method-dependent.

\begin{figure}
    \centering
    \includegraphics[width=\linewidth,angle=0]{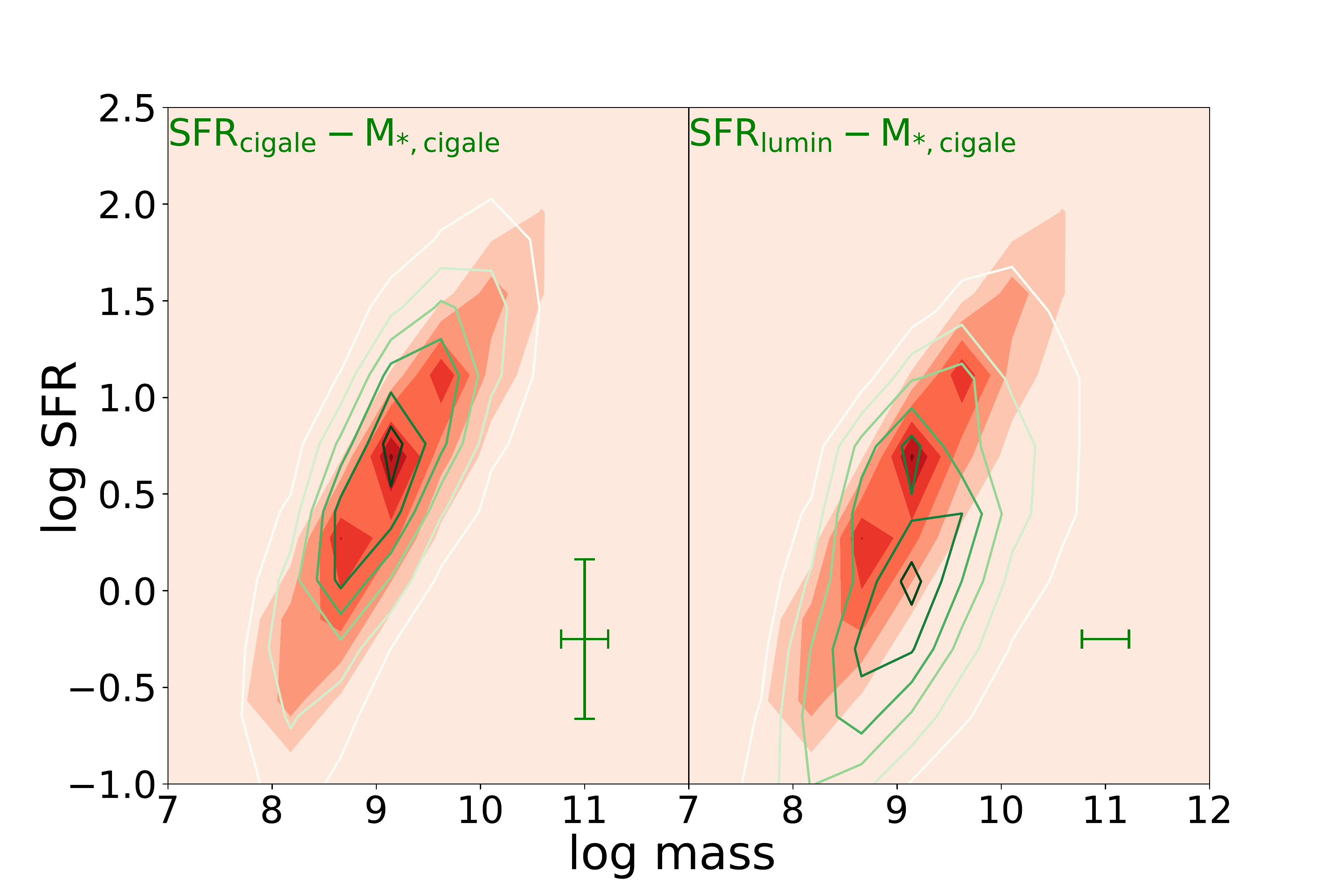}
    \caption{\small (Left) Comparison for 92,161 sources between our SFR$_{\rm cigale}$--$M_{\star,\rm cigale}$ (green open contours) and the \cite{speagle} MS relation (red filled contours). 
(Right) Same as the left panel, but for our SFR$_{\rm lumin}$--$M_{\star,\rm cigale}$.
In both panels, sources with SFR $<$0.1 $M_\odot$ yr$^{-1}$ are not included.}
    \label{fig:MS}
\end{figure}

\begin{figure}
    \centering
    \includegraphics[width=\linewidth,angle=0]{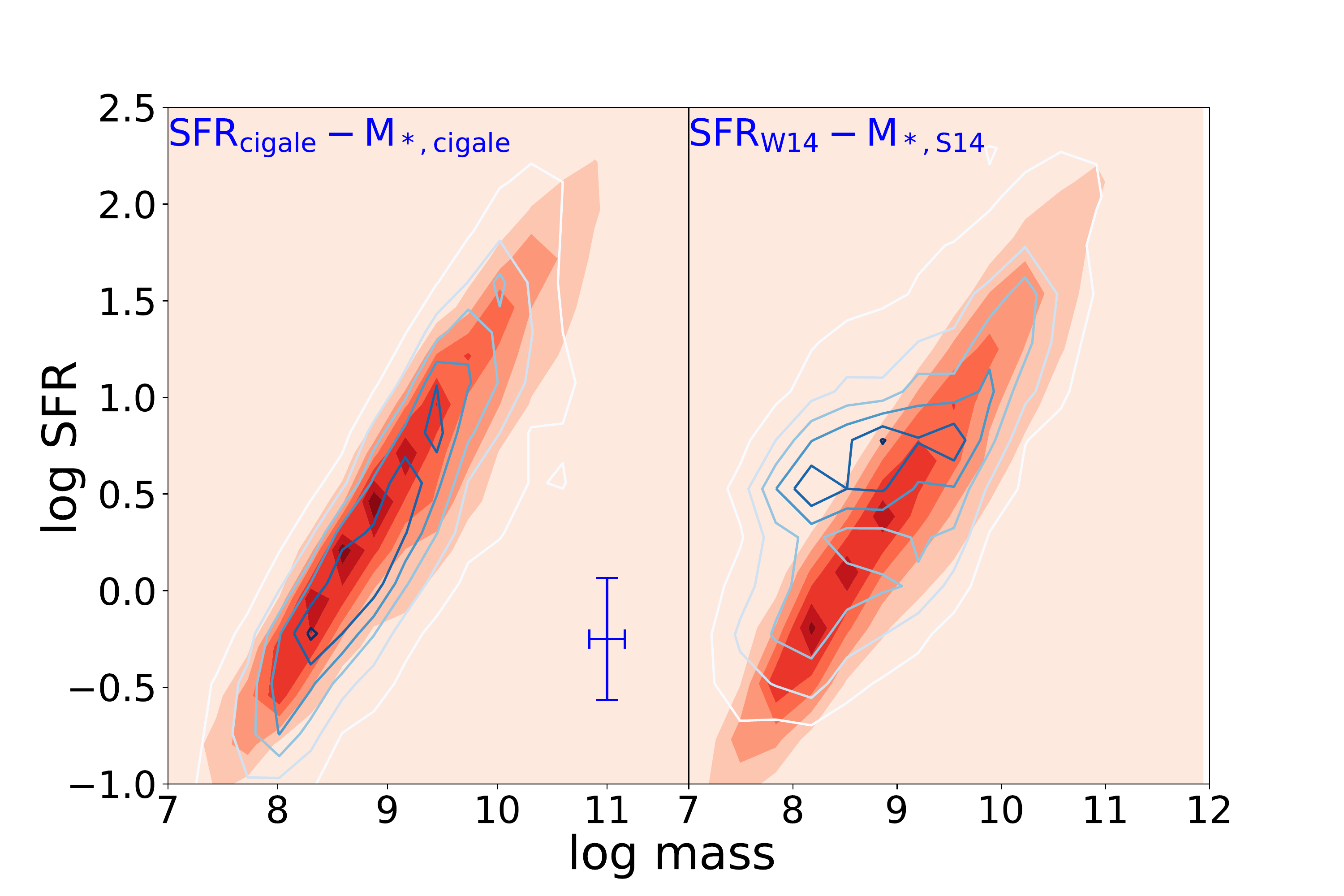}
    \caption{\small  (Left) Comparison for 8,918 sources between our SFR$_{\rm cigale}$--$M_{\star, \rm cigale}$ (blue open contours) and the \cite{speagle} MS relation (red filled contours). 
(Right) Same as the left panel, but for the W14 SFR plus S14 mass.
In both panels, sources with SFR $<$0.1 $M_\odot$ yr$^{-1}$ are not included.}
    \label{fig:MS_S14}
\end{figure}

As a sensible sanity check, in Figure \ref{fig:MS} we then directly compare our SFR$_{\rm cigale}$ and SFR$_{\rm lumin}$ estimates with those derived using the empirical star-forming main-sequence (MS) relation presented in \cite{speagle}. \cite{speagle} compiled 25 studies on MS from the literature to derive a robust functional form. We also plot a comparioson in Figure \ref{fig:MS_S14} including  8918 sources both in our work and W14 with SFR $>$0.1 $M_\odot$ yr$^{-1}$ to consolidate our results.

The red filled contours of both panels in Figure \ref{fig:MS} are obtained using our stellar-mass estimates in conjunction with the expected SFRs
according to the \cite{speagle} MS relation given our stellar-mass estimates, 
while the green open contours are for our stellar-mass and SFR$_{\rm cigale}$ (SFR$_{\rm lumin}$) estimates in the left (right) panel. For Figure \ref{fig:MS_S14}, the red filled contours are plotted using respective stellar mass estimates (ours in the left panel and S14 in the right panel) combined with their corresponding expected SFRs, while the blue open contours are made using mass and SFR etimates derived from SED fitting in our work (left) and from the S14 stellar mass plus W14 SFR (right).
Evidently, our SFR$_{\rm cigale}$--$M_{\star,\rm cigale}$ relation is in good agreement with the \cite{speagle} MS relation (left panels of Figures \ref{fig:MS} and \ref{fig:MS_S14}),
and there is a systematic offset and an apparent misalignment between the SFR$_{\rm lumin}$--$M_\star$, SFR$_{\rm W14}$--$M_\star$ relation and the \cite{speagle} MS relation (right panels of Figures \ref{fig:MS} and \ref{fig:MS_S14}). We argue that this misalignment results from two different methods used in deriving two parameters, i.e., calcultion from empirical relation and SED fitting. While for the left panels, two parameters are both derived from SED fitting and thus self-consistent.

Therefore, our SFR estimates are reliable given that we rely on an optimal combination of 
the most updated and accurate redshifts and PSF-matched photometry as well as appropriate parameter choices with the CIGALE code that is based on sophisticated Bayesian analysis, although it is challenging to obtain solid SFR estimates.

\section{Discussion}\label{sec:discussion}

\subsection{Influence of FIR Data}\label{subsec:fir_influence}

\begin{figure}[h]
    
    \centering
    \includegraphics[width=\linewidth,angle=0]{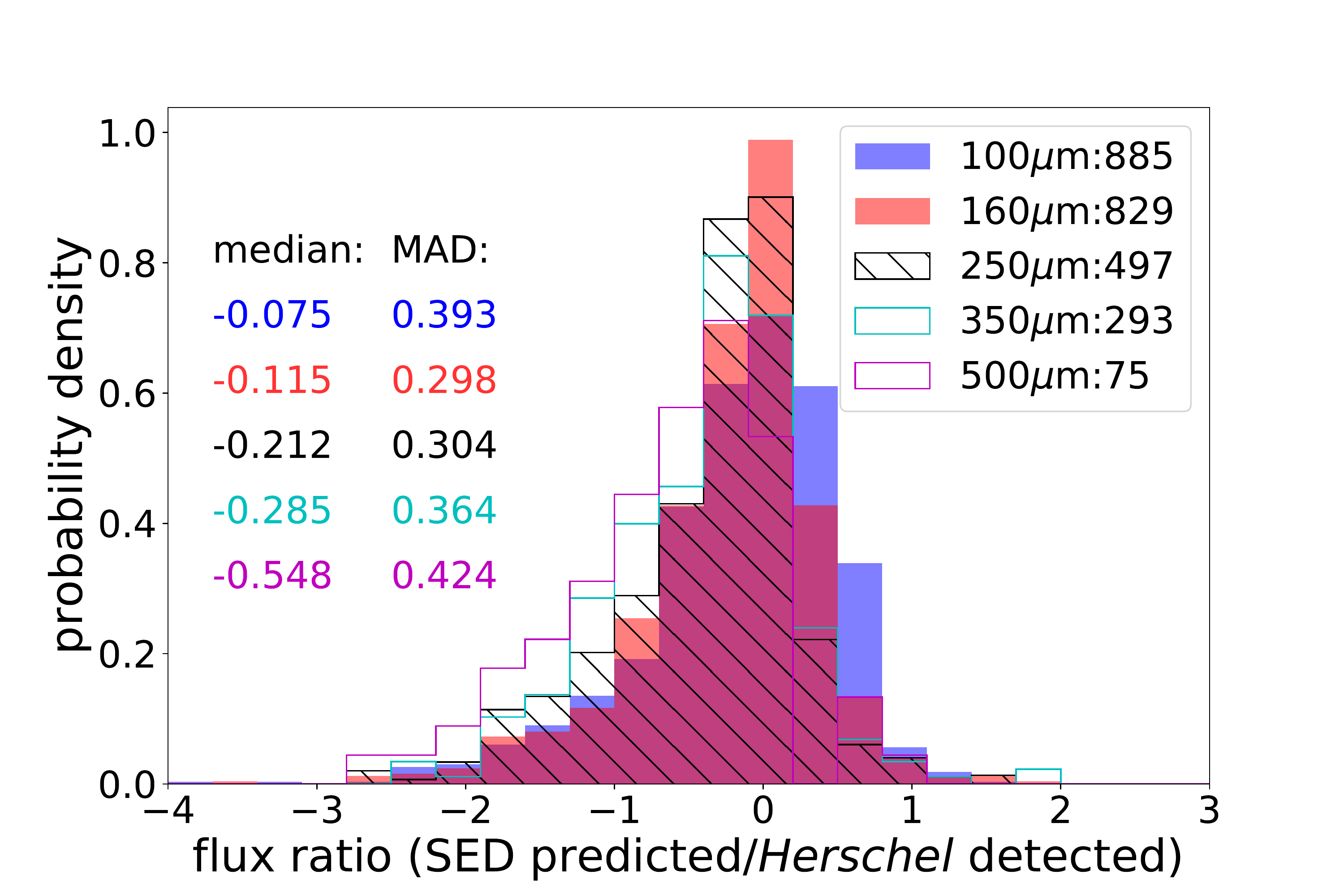}
    \caption{{\small Comparison between best-fit SED predicted fluxes with real detections at five \emph{Herschel} passbands. The numbers of sources available for comparison are indicated in the legend.}}
    \label{fig:herpd}
\end{figure}

Since a portion of the stellar emission at UV/optical wavelengths is absorbed and re-radiated in far-IR (FIR), the \emph{Herschel} PACS and SPIRE instruments, which cover the observed-frame 100, 160, 250, 350, and 500 $\rm \mu m$ wavelengths, can provide vital constraints on the galaxy FIR SED. 
However, the vast majority of our sources do not have any \emph{Herschel} detection/coverage, therefore the SFRs derived above do not utilize any \emph{Herschel} data. 
To check the reliability of our SED fitting result without including \emph{Herschel} data, we compare the predicted FIR fluxes from the best-fit SEDs with the observed \emph{Herschel} fluxes. 
We cross-match with the GOODS-\emph{Herschel} \citep{Elbaz} and \emph{Herschel}/PEP \citep{Lutz} catalogs to generate a subsample that contains 2205 sources with at least one MIPS 24 $\rm \mu m$ or \emph{Herschel} band detection, and compile their PACS and/or SPIRE fluxes. 
We then convolve the best-fit SEDs with the five PACS and SPIRE filter response curves to calculate the predicted fluxes and directly compare them with real detections. As shown in Figure \ref{fig:herpd}, the logarithmic ratios between the predicted PACS 100 $\rm \mu m$ and 160 $\rm \mu m$ fluxes to the observed ones peak close to zero (with median offsets $\approx$0.1 dex), indicating that, statistically, our best-fit SEDs are able to reproduce the observed emission at wavelengths extending to 160 $\rm \mu m$, albeit with MAD$\approx$0.3--0.4 dex. 
However, at 250, 350, and 500 $\rm \mu m$, we systematically underestimate the fluxes by $\approx$0.2--0.5 dex. We note that due to the low angular resolution in the SPIRE passbands (i.e., 18\arcsec~at 250 $\rm \mu m$, 25\arcsec~at 350 $\rm \mu m$, and 36\arcsec~at 500 $\rm \mu m$, respectively), the source confusion problem dramatically limits the accuracy of the measured photometry \citep{shu}. 
The SPIRE fluxes may be composed of contributions of several objects rather than an individual source, which can induce significant overestimation of the real fluxes. 
Therefore, the systematic difference between the predicted and observed fluxes at SPIRE wavelengths may be due to inaccurate photometry rather than any SED fitting issue, in contrast to
the consistency of fluxes at PACS wavelengths.

So far, the longest wavelength we adopt in the SED fitting is the IRAC 8 $\rm \mu m$, and only 58\% of our sources possess real detections in this band. The SFRs may be poorly constrained due to the lack of MIR and FIR data. 
To further check the reliability of the derived galaxy parameters without longer wavelength data, we perform a simple test. Unlike before, this time we include the MIR and FIR data (i.e., $\ge \rm 24\ \mu m$) in the fitting routine to obtain the best-fit SEDs. 
In the meantime, we carry out the SED fitting using the MIR and FIR data alone by minimizing the $\chi^2$ values between other galaxy templates and observed data. We then calculate the IR luminosities with these templates, and compare them to those calculated from SED fitting with and without MIR and FIR data respectively, in order to check whether the lack of longer wavelength data significantly biases our result. 

We utilize two sets of other galaxy templates. \citet[][hereafter CE01]{CE01} presented 105 MIR--FIR templates based on local galaxies. In addition, \citet[][hereafter K12]{K12} published two sets of SED templates for $z<1$ and $z>1$ star-forming galaxies. They combined deep photometry and Spitzer spectroscopy for 151 IR luminous galaxies in the GOODS-N and E-CDF-S fields, in an attempt to decompose the star-forming activity and AGN contribution in the MIR band. These templates are ideal for our studies since a prominent fraction of our sources are high-redshift star-forming galaxies. 

\begin{figure}
    \centering
    \includegraphics[width=\linewidth,angle=0]{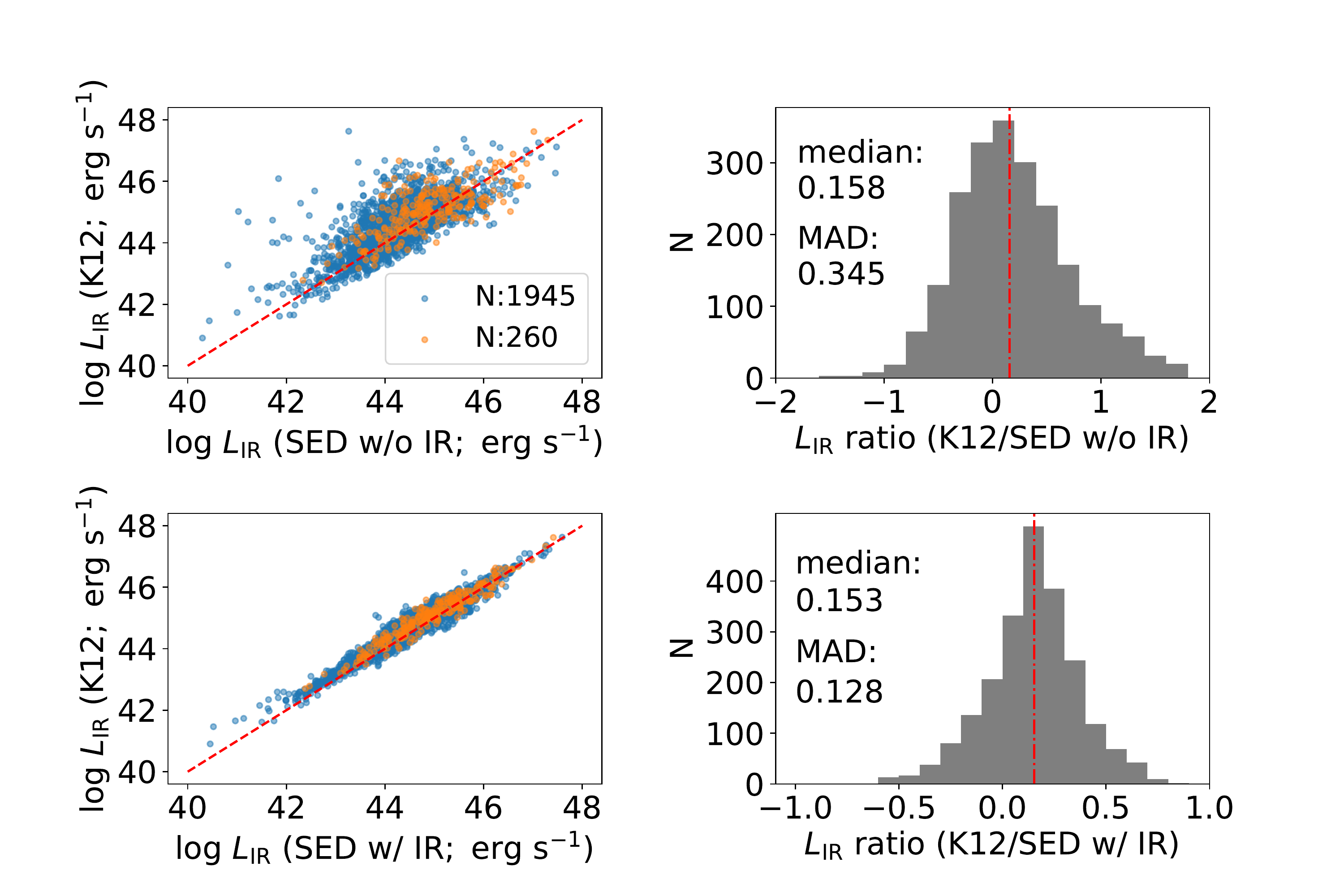}
    \caption{\small (Upper) Comparison between IR luminosities integrated using our best-fit templates (without MIR/FIR data; denoted as ``w/o IR'' hereafter) and best-fit K12 templates for 2205 sources. The red dashed line represents the 1-to-1 relation. (Lower) Comparison between IR luminosities integrated using our best-fit templates (with MIR/FIR data; denoted as ``w/ IR'' hereafter) and best-fit K12 templates. The orange and blue dots represent AGNs and normal galaxies respectively in the left panels, while the black histograms in the right panels do not separate them.}
    \label{fig:k12}
\end{figure}

\begin{figure}
    \centering
    \includegraphics[width=\linewidth,angle=0]{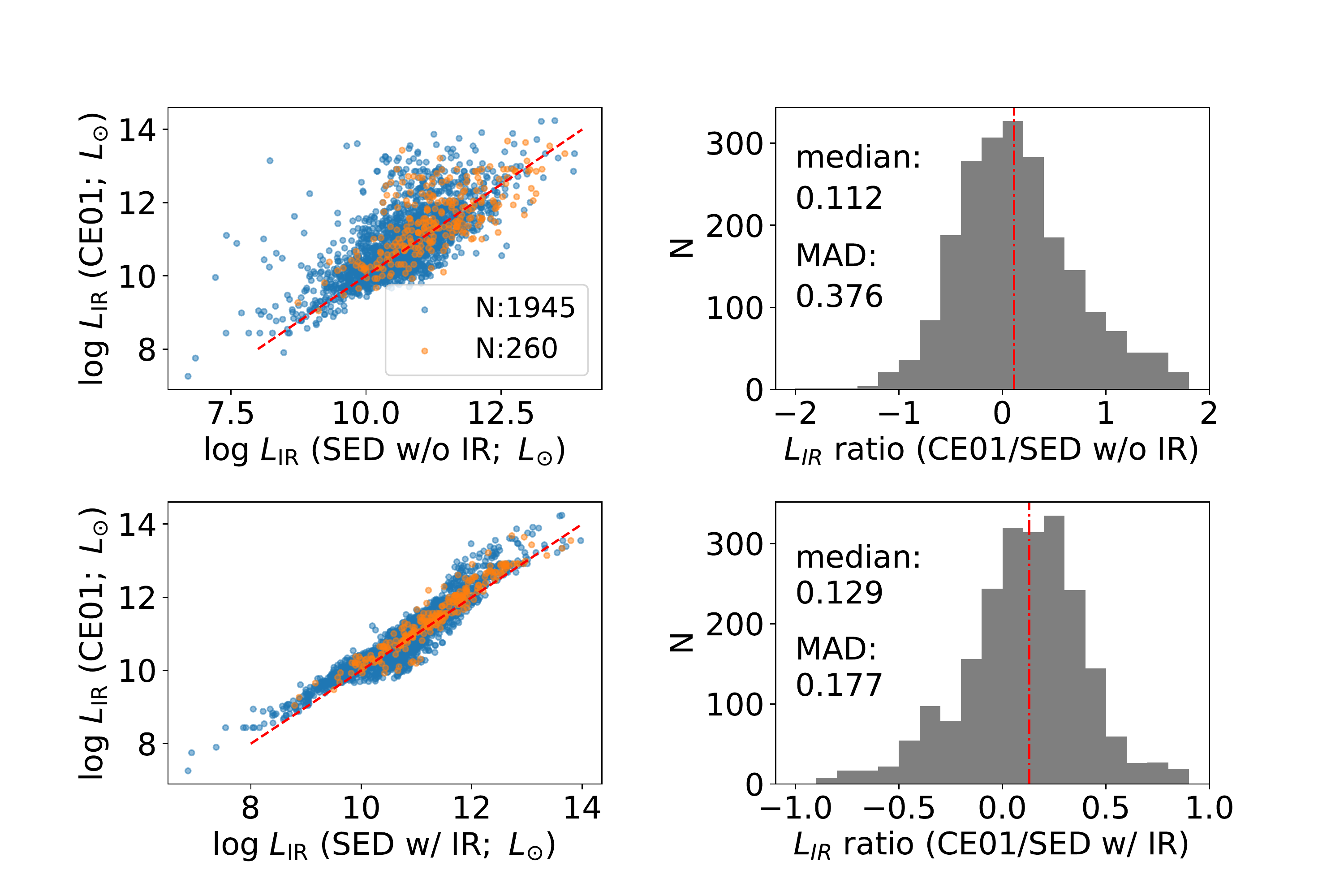}
    \caption{\small Same as Figure \ref{fig:k12}, but for comparison between using our best-fit templates (without and with MIR/FIR data) and best-fit CE01 templates.}
    \label{fig:ce01}
\end{figure}

We then convolve the K12 and CE01 templates with the filter transmission curves and compare them with observed MIR and FIR data points to determine the best normalizations by minimizing the $\chi^2$.  
Then we make direct comparison of the rest-frame 8-1000 $\mu$m IR luminosities with the values obtained through broadband SED fitting under two circumstances (i.e., with and without MIR--FIR data). The results of fitting K12 and CE01 templates are shown in Figure \ref{fig:k12} and Figure \ref{fig:ce01}, respectively. 
The upper (lower) panels are the IR luminosity comparison between the cases of using the K12/CE01 templates with only the MIR and FIR data and performing broadband SED fitting without (with) MIR and FIR data. 
It seems that the two sets of IR luminosities are more tightly correlated (i.e., smaller MAD) when MIR--FIR data being involved, in contrast to the larger scatters when MIR--FIR data not being used. We compare AGNs (see Section \ref{subsec:agn}) with normal galaxies, and find that they share the same trend, i.e., with similar distributions, median and MAD values.
This indicates that the inclusion of MIR and FIR data does affect the fitting accuracy (i.e., in terms of MAD values) but not the median offsets, no matter whichever the source is, a normal galaxy or an AGN. 

\begin{figure}
    \centering
    \includegraphics[width=\linewidth]{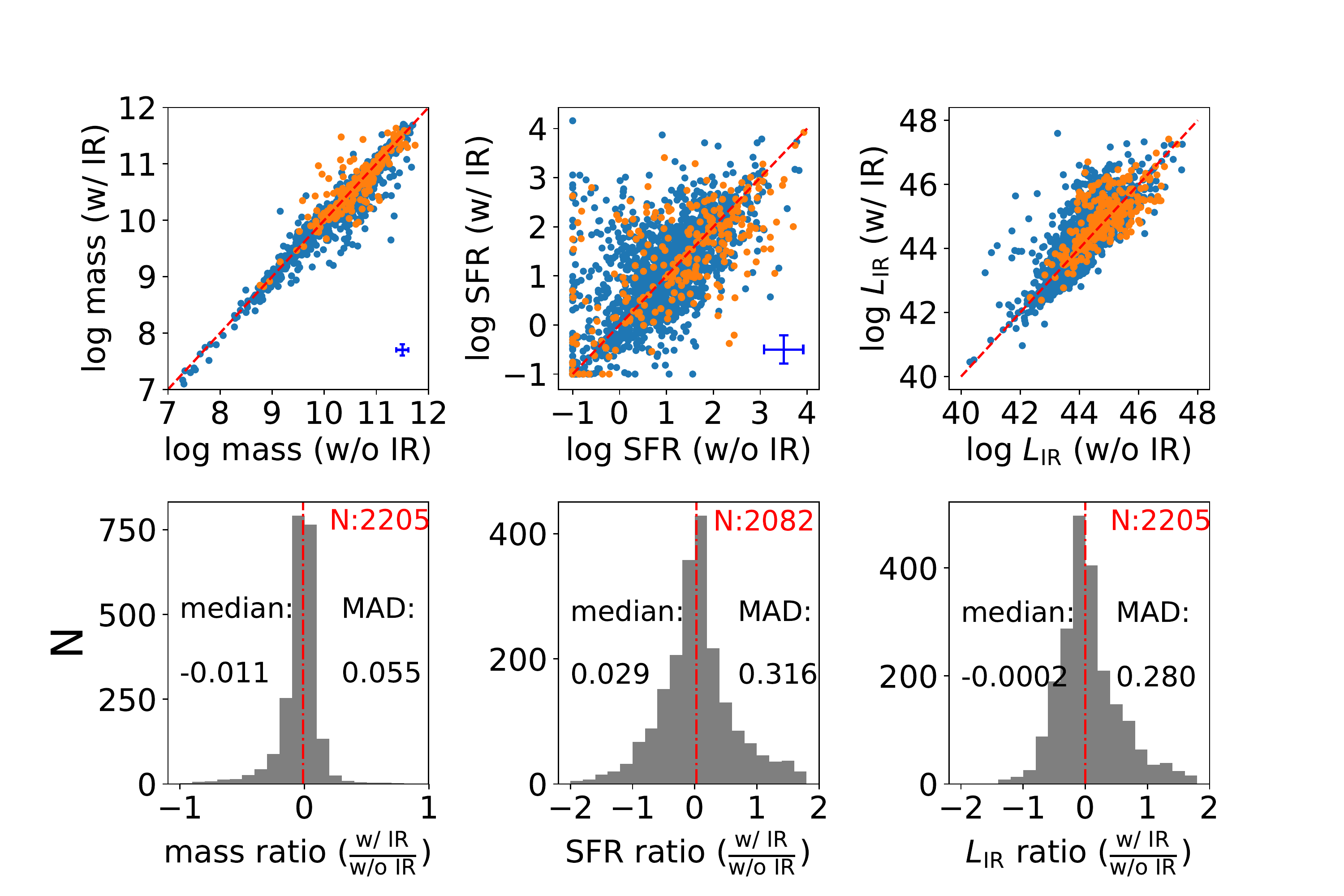}
    \caption{\small Comparisons of stellar mass (Left), SFR (Middle; sources with $<0.1\ M_\odot$~yr$^{-1}$ not included), and IR luminosity (Right) estimates between the cases of fitting with and without MIR--FIR data. The orange and blue dots represent AGNs and normal galaxies respectively in the top panels, while the black histograms in the bottom panels do not separate them.}
    \label{fig:wwoirlir}
\end{figure}

Then we directly compare the results of the 2205 CIGALE derived stellar masses, SFRs, and IR luminosities (using K12 templates) with and without MIR--FIR data, as shown in Figure \ref{fig:wwoirlir}. 
The left panels show that the two sets of stellar-mass estimates are in excellent agreement with each other. This is expected because the absence of MIR and/or FIR data should not have significant impact on the stellar-mass estimation since it is mainly constrained from the rest-frame optical colors that are well-covered by our broadband SEDs. 
In contrast, the SFR (i.e., the middle panels) and IR luminosity (i.e., the right panels) estimates under two circumstances show larger scatters (yet no systematic offsets), which demonstrates again that MIR and FIR data do affect the accuracy of SFR estimation.

In conclusion, although our SED fitting can predict \emph{Herschel} PACS fluxes even when no MIR/FIR data are included, it only works for monochromatic fluxes up to 160$\mu$m with large scatters. 
When it comes to IR luminosity and SFR estimates, the ratios between the results with and without MIR/FIR data show no apparent systematic offsets but large scatters, indicating that the accuracy of SFR estimation is highly (MIR/FIR) data-dependent.

\subsection{Influence of SFH Models}\label{subsec:sfh}

Previous studies have shown that both the delayed-$\tau$ model adopted in this work and the widely-used 2$\tau$-dec model can well reproduce $M_*$ and SFRs by fitting the mock galaxy SEDs \citep{ciesla15,ciesla16}. They also found that the 2$\tau$-dec model may have better performance on recovering the input $M_*$ and SFRs, although it has an issue in calculating the galaxy age since it significantly underestimates this parameter. 
As we discussed in Section \ref{subsec:sfr}, unlike stellar-mass estimation, the SFR estimates using the delayed-$\tau$ model are relatively less constrained and suffer large dispersion when compared with previous works. 
To test whether the 2$\tau$-dec model can better constrain $M_*$ and SFRs using real observed data, we change our SFH into the 2$\tau$-dec model and re-fit the galaxy SEDs. 
The 2$\tau$-dec model divides stars into old and young populations and is expressed by adding a late burst to the 1$\tau$-dec SFH: 
\begin{eqnarray}  
\rm SFR(t)=     
\left\{                  
\begin{array}{lll}     
e^{-t/\tau_1},\ t\ <  t_1 - t_2\\  
e^{-t/\tau_1} + k \times e^{-t/\tau_2},\ t\ > t_1 - t_2,\\  
\end{array}           
\right.  
\label{eq:tau2}            
\end{eqnarray}
where $\tau_1$ and $\tau_2$ ($t_1$ and $t_2$) are the e-folding times (ages) of the old and young stellar populations, respectively; 
and $k$ is the mass fraction of the late burst. 
The adopted parameter space for the 2$\tau$-dec model is listed in Table \ref{table:t6}. The results are directly compared with those obtained through the delayed-$\tau$ model as well as the reference works. We note that here we do not take other SFH models such as constant SFH, linearly increasing SFH, or truncated SFH into consideration, since the galaxy parameters derived using these models deviate severely from the 1$\tau$-dec and delayed-$\tau$ models as shown in \cite{S15}.

\begin{figure}
    \centering
    \includegraphics[width=\linewidth,angle=0]{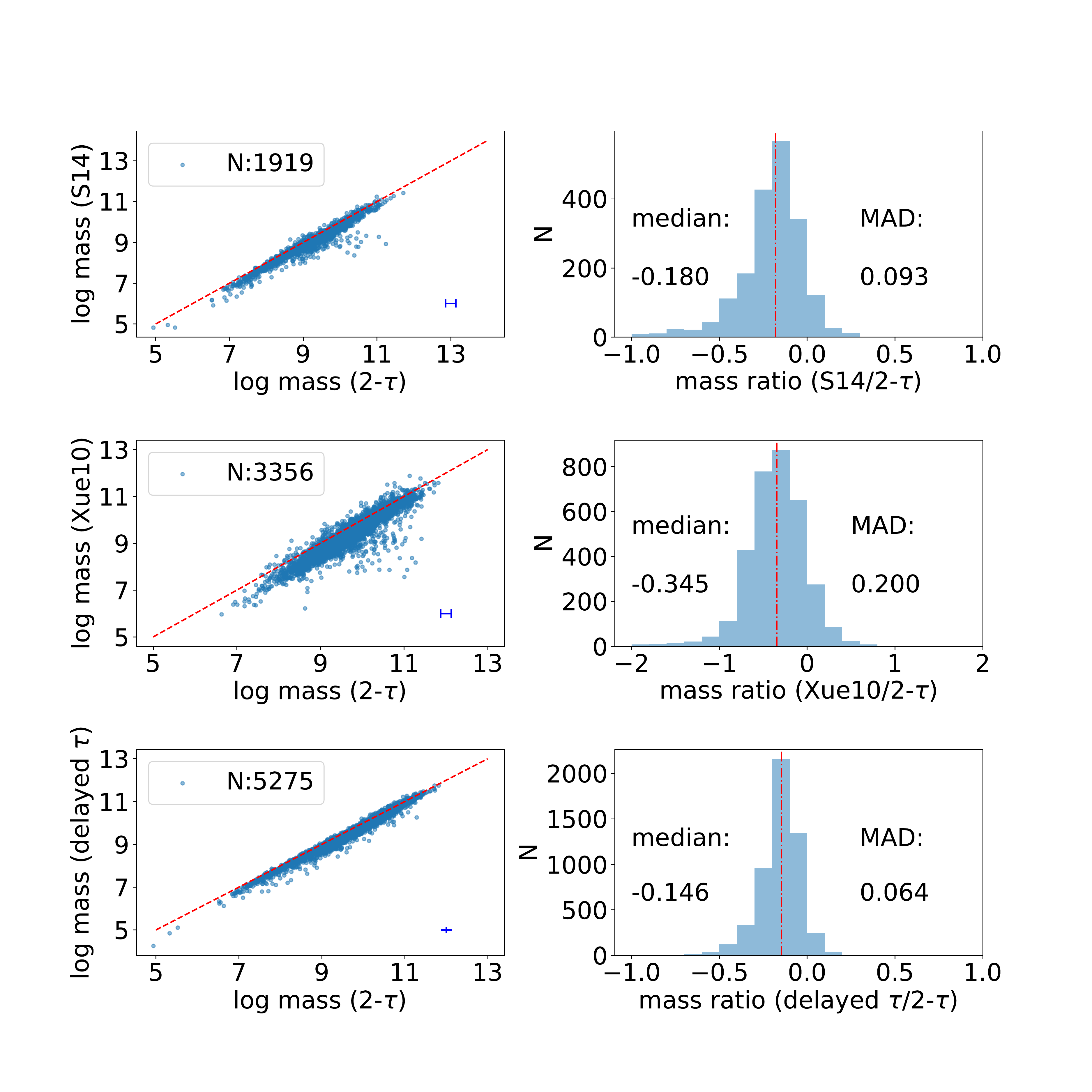}
    \caption{\small Comparisons between the S14 (Top row), Xue10 (Middle row), and our delayed-$\tau$ derived (Bottom row) stellar-mass estimates and those derived with the 2$\tau$-dec SFH model, respectively.}
    \label{fig:sfhnew1}
\end{figure}

\begin{figure}
    \centering
    \includegraphics[width=\linewidth,angle=0]{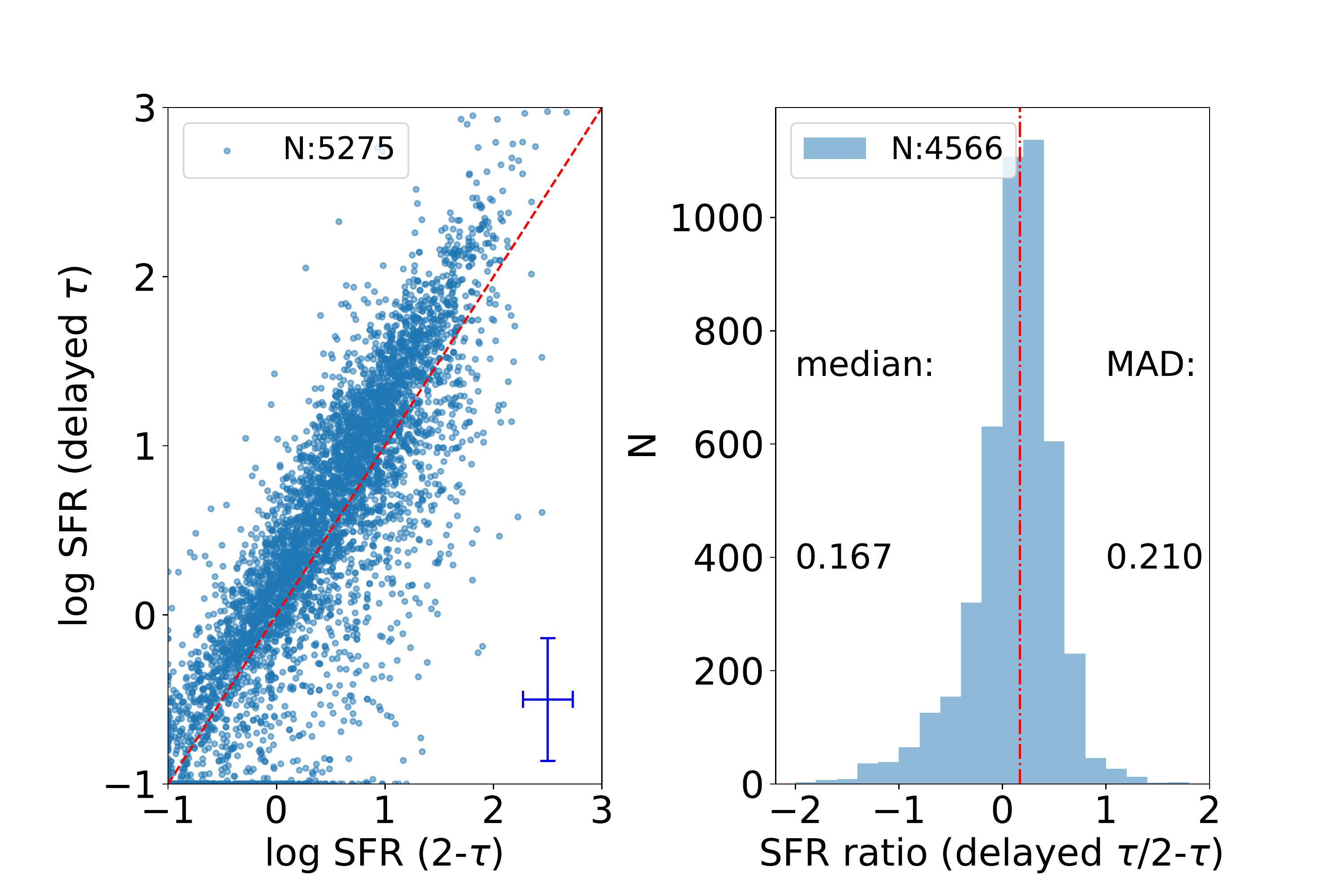}
    \caption{\small Comparison between our delayed-$\tau$ derived SFR estimates and those derived with the 2$\tau$-dec SFH model. In the right panel, sources with either SFR$\le$0.1 $M_\odot$ yr$^{-1}$ are not included.}
    \label{fig:sfhnew2}
\end{figure}

The size of the template library under the 2$\tau$-dec model is dramatically larger than that under the delayed-$\tau$ model due to its additional parameters. 
We only consider sources with better photometry (e.g., having data from $=$25 filters for the comparison between the S14 and 2$\tau$-dec results) and adopt simple parameter space (see Table \ref{table:t6}) to avoid memory crash and speed up the computation. 
Figure \ref{fig:sfhnew1} presents three sets of comparisons between the S14, Xue10, and our delayed-$\tau$ stellar-mass estimates and the 2$\tau$-dec ones, respectively.
It is clear that, for all these comparisons, the 2$\tau$-dec stellar masses are systematically larger than the other estimates, with median offsets ranging from $-0.146$ to $-0.345$ dex and MAD$\approx$0.1--0.2 dex.
Figure \ref{fig:sfhnew2} shows the comparison between our delayed-$\tau$ SFR estimates and the 2$\tau$-dec ones, indicating that
the 2$\tau$-dec SFRs are systematically smaller than the delayed-$\tau$ SFRs, with a median offset of $0.167$ dex and MAD=0.210 dex.
Given the above analysis, we favor the results derived from the delayed-$\tau$ SFH model.

\begin{table}
\begin{center}
\caption[]{Parameter Space of the 2$\tau$-dec Model}\label{table:t6}
\begin{tabular}{ccc}
\hline\noalign{\smallskip}
Module & Parameter & Value \\\hline
star-formation history & $\tau_1$ of the old population& 1, 10, 100, 1000, 10000 Myr \\
2$\tau$-dec model &  $\tau_2$ of the young population & 1, 10, 100, 1000, 10000 Myr  \\
& age ($t_1$) of the old population &1000, 5000, 10000 Myr\\
& age ($t_2$) of the young population &1, 10, 100, 1000Myr\\
&burst fraction ($k$) &0.01, 0.05, 0.1 \\
\noalign{\smallskip}\hline
    dust attenuation &E$(B-V)$ of young population & 0--1 (in steps of 0.2)\\
  \noalign{\smallskip}\hline
\end{tabular}
\end{center}
\end{table}

\subsection{Influence of AGN Contribution}\label{subsec:agn}

\begin{figure}
    \centering
    \includegraphics[width=\linewidth,angle=0]{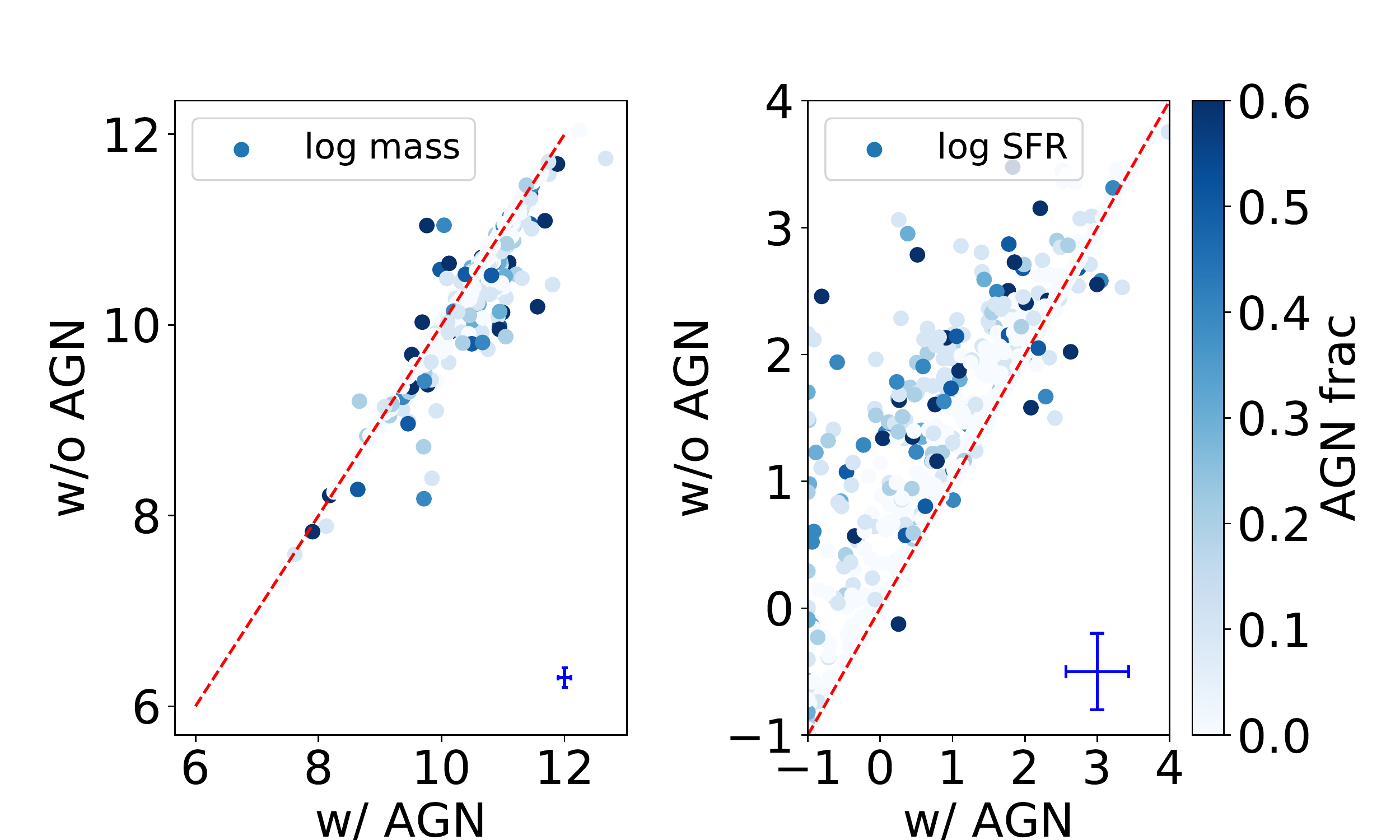}
    \caption{\small Comparisons of stellar mass (Left) and SFR (Right) estimates between the cases of SED fitting with and without AGN contribution. The data points are color-coded according to their respective AGN fractions.}
    \label{fig:agnfrac}
\end{figure}

During the fitting routine, we assign the AGN fraction in the \cite{Dale} model a series of values from 0.0 to 0.6 (see Table \ref{table:t5}). 
In order to check whether the inclusion of AGN contribution makes a difference, we use the 576 X-ray AGNs (see Table \ref{table:t3}) as a subsample and make comparisons of the fitting derived values with those derived without contribution (i.e., fixed at 0) from AGNs. 
The result is shown in Figure \ref{fig:agnfrac}. 
It is evident that stellar-mass estimates are roughly distributed along the 1-to-1 line, in spite of an over-density occurring under the 1-to-1 line. 
This over-density is caused by that 
taking AGN contribution into account results in a slightly redder stellar continuum emission for the galaxy component, which eventually leads to a best fit of an older stellar population and thus a higher stellar mass. 
In contrast, SFR estimates in the case of AGN contribution being considered are systematically smaller than those not considering AGN contribution, especially for those with large AGN contributions, due to a portion of UV and IR emission being accounted for by AGNs. 
Therefore, we make sure that this parameter of AGN contribution does affect the final results. 
For the sources not detected in the 2 Ms CDF-N \citep{Xue16}, we still invoke this varying parameter since some weak and/or highly obscured AGNs may be missed in the \emph{Chandra} catalog. Even for these galaxies, it is still beneficial to decompose the contribution from AGN and star-forming activities in order to obtain more reliable estimates of galaxy properties.

\section{Final Catalog}\label{sec:cat}

We provide a final catalog of source properties and make it publicly available. For easy use of our catalog, in Table \ref{table:t7} we present our catalog that consists of a total of 36 columns for the 178,341 sources (for completeness, we also include stars and sources without redshift information), and describe their details below:

1. Column 1 gives the source identity number (id) ranging from 1 to 178,341. Sources are sorted according to their appearance orders in the three base catalogs \citep[in the order of Y14, S14 and][]{Wang10}.

2. Columns 2 and 3 give the right ascension (RA) and declination (DEC) in the J2000.0 frame (in units of degree), respectively. The positions adopted from S14 and \cite{Wang10} are corrected by the mean deviations calculated in the first cross-match step as mentioned in Section \ref{sec:data} (i.e., RA$_{\rm S14}$=RA$_{\rm S14,original}-0.147$\arcsec; DEC$_{\rm S14}$=DEC$_{\rm S14,original}-0.071$\arcsec; RA$_{\rm Wang10}$=RA$_{\rm Wang10,original}-0.207$\arcsec; DEC$_{\rm Wang10}$=DEC$_{\rm Wang10,original}-0.086$\arcsec).

3. Column 4 gives the preferred redshift value, Column 5 gives the type information, and Column 6 lists the corresponding reference index rz (see Table \ref{table:t2}). The white dwarfs (24 sources; see Y14) and stars (4,678 sources) are represented by $-2$ and $-1$, respectively. Galaxies (122,514 sources) are labeled as 0 while X-ray AGNs (576 sources) are marked as 1.
Sources that are too faint to be distinguished in S14 (24,961 sources) are labeled as 2. The remaining sources do not have a type classification in the original catalogs.
When no information is available, the type classification and reference are set to the default value $-99$.

4. Column 7 gives the number of filters (including upper limits) used in the SED fitting.
The minimum, maximum, mean, and median numbers of filters are 2, 25, 12.6, and 13, respectively. We caution that for sources with a small number of filters, care should be taken when using the fitting results.

 5. Columns 8-11 give 3 flags extracted from Y14 and one flag extracted from S14, respectively. Qz flag indicates the photometric redshift quality where lower values represent higher quality. S/N ratio flag gives the source-detection significance. Cr flag indicates whether the source is in the central H-HDF-N region. Column 11 is from S14 that indicates the quality of photometry (for more details, refer to section 3.8 in S14). These flags serve as slection criteria in addition to filter number (column 7) and reduced $\chi^2$ (column 12) when using our fitting results.

6. Columns 12--36 give the SED-fitting results for the 145,635 sources that are classified/treated as galaxies and have redshift information available (the remaining sources have all these columns set as ``$-$''), including the reduced $\chi^2$, stellar mass and associated uncertainty (logarithmic scale; in units of M$_{\odot}$), SFR and associated uncertainty (logarithmic scale; in units of M$_{\odot}/$yr; any value of $<0.1$ set to 0.1 M$_{\odot}/$yr), characteristic timescale $\tau$ (Myr) and associated uncertainty, age of the oldest population (Myr) and associated uncertainty, metallicity and associated uncertainty, attenuation $A_{\rm V}$ (mag) of old and young populations and their associated uncertainties, AGN fraction, and 9-band flux densities derived from the best-fit SED (i.e., \emph{U}, \emph{B}, \emph{V}, \emph{R}, \emph{I}, \emph{z}, \emph{J}, \emph{H}, and \emph{Ks} band; mJy).





\begin{table}
	\begin{center}
		\caption[]{Final Catalog}\label{table:t7}
		\begin{tabular}{cccc}
			\hline\noalign{\smallskip}
			Index&Column &Unit& Meaning \\\hline
			1&id&---&sequence number \\2&RAdeg&degree& right ascension in J2000 frame\\3&DEdeg&degree&declination in J2000 frame \\
			4&redshift&---& spectroscopic or photometric redshift \\5&type&---&type information \\6&rz&---& reference of redshift \\
			7&filter&---&number of filters (including upper limits) \\
			8&Qz\_Y14&---&redshift quality parameter extracted from Y14 \\9&S/N\_Y14&---&S/N ratio extracted from Y14 \\10&cr\_Y14&---&flag extracted from Y14 \\11&use\_phot\_S14&---&flag extracted from S14 \\
			12&red\_chi2&---&reduced $\chi^2$ \\13&mass&M$_{\odot}$&mass in logarithmic unit\\14&mass\_err&M$_{\odot}$&mass uncertainty in logarithmic unit \\
			15&SFR&M$_{\odot}$/yr&SFR in logarithmic unit \\16&SFR\_err&M$_{\odot}$/yr&SFR uncertainty in logarithmic unit \\
			17&$\tau$ & Myr &characteristic timescale \\18&$\tau$\_err&Myr&uncertainty of characteristic timescale \\
			19&age&Myr&age of the oldest population \\20&age\_err&Myr&uncertainty of age of the oldest population\\
			21&metallicity&---&metallicity \\22&metallicity\_err&---&uncertainty of metallicity \\
			23&AV\_old&mag&AV of old population\\24&AV\_old\_err&mag&uncertainty of AV of old population\\
			25&AV\_young&mag&AV of young population\\26&AV\_young\_err&mag&uncertainty of AV of young population \\
			27&AGNfrac&---&AGN fraction defined in \cite{Dale} \\28-36&U-Ks&mJy&flux density in best-fit SED\\
			\noalign{\smallskip}\hline
		\end{tabular}
	\end{center}
	\tablecomments{0.96\textwidth}{The full table contains 36 columns of information for 178,341 sources, which is available in the online journal.}
\end{table}

\section{Summary}\label{sec:sum}

In this work, we compile multi-wavelength photometry ranging from UV to IR and other valuable information such as redshift and type classification for a large sample of 178,341 sources in the H-HDF-N field (see Tables~\ref{table:t1}, \ref{table:t2}, and \ref{table:t4}).
Taking into account influences of MIR/FIR data, SFH models, and AGN contribution (see Section~\ref{sec:discussion}), we utilize the SED-fitting code CIGALE that is based on Bayesian analysis to derive stellar masses, SFRs, and other properties for 145,635 sources among the full sample, which are classified/treated as galaxies and have redshift information available (see Section~\ref{sec:sed} and Table~\ref{table:t3}). 

Through various consistency and robustness check (see Section~\ref{subsec:mass}), we find that our stellar-mass estimates are in excellent agreement (i.e., a median offset of $<$0.1 dex and MAD $\sim$0.17 dex) with that of previous works, which involved different photometry and were derived from SED fitting with the FAST code (S14) or based on the mass-to-light ratios (Xue10). Therefore we conclude that stellar mass is a parameter that can be robustly retrieved via broadband SED fitting. 

In contrast, 
the estimation of SFR is more challenging. 
Unlike stellar-mass estimation, SFR estimation
depends much more sensitively on the calculation method (see Section \ref{subsec:sfr}), the wavelength coverage of the photometric data (especially the \emph{Herschel} FIR data; see Section \ref{subsec:fir_influence}), the assumed SFH (see Section \ref{subsec:sfh}), and the likely AGN contribution (see Section~\ref{subsec:agn}). 
Nevertheless, we manage to obtain reliable SFR estimates (see Section \ref{subsec:sfr}), as we adopt
the most updated and accurate redshifts and PSF-matched photometry (see Section~\ref{sec:data}) and make sensible
parameter choices with the CIGALE code (see Section~\ref{sec:sed}).

We make our catalog of galaxy properties (including, e.g., redshift, stellar mass, SFR, age, metallicity, dust attenuation; see Section~\ref{sec:cat} and Table \ref{table:t7}) publicly available, which is the largest of its kind in the H-HDF-N. This catalog should be beneficial to many future studies that aim at improving our understanding of various questions regarding the formation and evolution of galaxies.

\noindent {\bf Acknowledgments}
We thank the referee for helpful comments that helped improve this paper. We acknowledge support from the 973 Program (2015CB857004), the National Natural Science Foundation of China (NSFC-11473026, 11421303), and the CAS Frontier Science Key Research Program (QYZDJ-SSW-SLH006).

\bibliography{ms}

\label{lastpage}

\end{document}